\documentclass[apj]{emulateapj}
\usepackage{apjfonts}
\usepackage{pdflscape}
\usepackage{afterpage}
\shorttitle{VHE $\gamma$-Ray Observation of the Crab Nebula and
Pulsar with MAGIC}

\shortauthors{Albert et al.}

\begin{document}


\title{VHE $\gamma$-Ray Observation of the Crab Nebula and its Pulsar with the MAGIC Telescope}


\author{
 J.~Albert\altaffilmark{a},
 E.~Aliu\altaffilmark{b},
 H.~Anderhub\altaffilmark{c},
 P.~Antoranz\altaffilmark{d},
 A.~Armada\altaffilmark{b},
 C.~Baixeras\altaffilmark{e},
 J.~A.~Barrio\altaffilmark{d},
 H.~Bartko\altaffilmark{f},
 D.~Bastieri\altaffilmark{g},
 J.~K.~Becker\altaffilmark{h}
% ...
}
 \altaffiltext{a} {Universit\"at W\"urzburg, D-97074 W\"urzburg, Germany}
 \altaffiltext{b} {Institut de F\'\i sica d'Altes Energies, Edifici Cn., E-08193 Bellaterra (Barcelona), Spain}
 \altaffiltext{c} {ETH Zurich, CH-8093 Switzerland}
 \altaffiltext{d} {Universidad Complutense, E-28040 Madrid, Spain}
 \altaffiltext{e} {Universitat Aut\`onoma de Barcelona, E-08193 Bellaterra, Spain}
 \altaffiltext{f} {Max-Planck-Institut f\"ur Physik, D-80805 M\"unchen, Germany}
 \altaffiltext{g} {Universit\`a di Padova and INFN, I-35131 Padova, Italy}
 \altaffiltext{h} {Universit\"at Dortmund, D-44227 Dortmund, Germany}
% ...
\submitted{Received 2007 May 18; accepted 2007 October 31}
\journalinfo{The Astrophysical Journal, 674:000-000, 2008
February 20}



\begin{abstract}
 We report about very high energy (VHE) $\gamma$-ray observations of
 the Crab Nebula with the MAGIC
 telescope. The $\gamma$-ray flux from the nebula was measured  between
 60\,GeV and 9\,TeV. The energy spectrum can be described by a
 curved power law ${\mathrm{d}F}/{\mathrm{d}E}=f_0\,
 \left(E/300\,\mathrm{GeV}\right)^{\left[a+b\log_{10}\left(E/300\,\mathrm{GeV}\right)\right]}
 $
 with a flux normalization $f_0$ of
 $(6.0\pm0.2_{\mathrm{stat}})\times10^{-10}\,$cm$^{-2}$s$^{-1}$TeV$^{-1}$,
 $a=-2.31\pm0.06_{\mathrm{stat}}$ and
 $b=-0.26\pm0.07_{\mathrm{stat}}$. The peak in the spectral energy distribution is estimated at
 $77\pm35\,$GeV. Within the observation time and the experimental
 resolution of the telescope, the $\gamma$-ray emission is steady and pointlike.
 The emission's center of gravity
 coincides with the position of the pulsar. Pulsed $\gamma$-ray emission from the pulsar could not be detected.
 We constrain the cutoff energy of the pulsed spectrum to
 be less than 27\,GeV, assuming that the differential
 energy spectrum has an exponential cutoff. For a super-exponential shape, the cutoff
 energy can be as high as 60\,GeV.
\end{abstract}





\section{Introduction}

The Crab Nebula is the remnant of a supernova explosion that
occurred in AD 1054 \citep[e.g.][and references
therein]{1999PASP..111..871C} at a distance of $\sim2\,$kpc. It is
one of the best-studied non-thermal celestial objects in almost
all wavelength bands of the electromagnetic spectrum from
$10^{-5}\,$eV (radio) to nearly $10^{14}$\,eV ($\gamma$-rays).
There is little doubt that the engine of the nebula is the pulsar
PSR B0531+21 (hereafter Crab pulsar) in its center.

In very high energy (VHE) $\gamma$-ray astronomy
%\footnote{We refer
%to VHE as $\gamma$-rays with energies between 10\,GeV and
%100\,TeV}
the Crab Nebula was first detected with large significance at TeV
energies by the pioneering Whipple telescope
\citep{1989ApJ...342..379W}. The nebula turned out to be the
strongest source of steady VHE $\gamma$-ray emission in the
Galaxy. It is therefore used as the standard ``calibration
candle'' for ground-based $\gamma$-ray experiments. Apart from
testing the performance of $\gamma$-ray instruments another aim of
measuring the Crab Nebula is to increase the measurement precision
of the Crab Nebula flux and the energy range covered. These
continuous efforts provide insights necessary for the
understanding of the very details of the emission mechanisms of
VHE $\gamma$-rays in the Crab Nebula and pulsar. Important
questions remain to be answered concerning the emission mechanisms
of the nebula, of the pulsar at GeV energies, and of the nebula
around PeV energies. Since its discovery, detailed studies of the
VHE emission energy spectrum, ranging from several hundred GeV up
to 80 TeV, have been carried out
\citep[e.g.][]{1990ICRC....2..135A,1991ApJ...377..467V,
1991ICRC....1..220B,1993A&A...270..401G,1996APh.....4..199H,1998ApJ...492L..33T,1998ApJ...503..744H,1999ApJ...525L..93A,
2002majumdar, 2004ApJ...614..897A,2006A&A...457..899A}. Between
10 and $\sim200\,$GeV, observations are sparse. A few
results are provided by converted solar concentrator arrays that
use the wave front sampling
technique~\citep{2001ApJ...547..949O,2002ApJ...566..343D,2002APh....17..293A,2006A&A...459..453S}.
However, the wave front sampling suffers from relatively large
uncertainties in the calculation of the effective area and of the
energy, as well as from poor $\gamma$/hadron discrimination,
which make it difficult to perform differential flux measurements.

A good explanation of the nebular dynamics and the observed energy
spectrum below GeV energies can be obtained with the
magneto-hydrodynamic model suggested first by
\cite{1974MNRAS.167....1R} and  developed further by
\cite{1984ApJ...283..694K,1984ApJ...283..710K}. In this framework,
the pulsar provides a continuous flow of charged particles (pulsar
wind) with Lorentz factors of $10^6-10^7$. A standing reverse
shock forms where the wind ram pressure balances the total
pressure of the nebula. Wind particles accelerate in the shock to
ultra-relativistic energies and subsequently lose their energy by
synchrotron emission. The presence of  synchrotron emission up to
a few hundred MeV in conjunction with the observed $\gamma$-ray
spectrum at TeV energies shows that particle acceleration takes
place up to energies of $\sim10^{15}-10^{16}\,$eV. From the total
luminosity of the synchrotron emission, it can be inferred that
about 10\% of the pulsar's energy loss rate is converted into the
kinetic energy of particles.

Above 1\,GeV  the dominant source of $\gamma$-ray emission is most
likely inverse Compton (IC) scattering of synchrotron photons by
the synchrotron-emitting electrons  in the shocked wind region
\citep[synchrotron self-Compton model \mbox{[}SSC\mbox{]};][]{1965PhRvL..15..577G,1989ApJ...342..379W,1992ApJ...396..161D}.
To explain the observed VHE $\gamma$-ray spectrum, several other
seed photon fields are also believed to contribute to the inverse
Compton scattering, namely, far-infrared excess, cosmic microwave
background, and millimeter-photons~\citep[e.g.][]{2004ApJ...614..897A}.

Although the IC-mechanism gives a good description of the observed
energy spectrum between 500 GeV and about 10 TeV, other processes
may  contribute in part  to the VHE $\gamma$-ray emission. It is
likely that a significant fraction of the mechanical energy lost
by the pulsar is taken away by a hadronic component in the wind.
Following interactions of this component with the interstellar
medium, VHE $\gamma$-rays are emitted by decaying $\pi^{0}$s,
which modify the energy spectrum at TeV energies and beyond
\citep{1996MNRAS.278..525A,1997PhRvL..79.2616B,2003A&A...405..689B,2003A&A...402..827A}.
\cite{1996MNRAS.278..525A} discuss the possibility of an
``amplified'' bremsstrahlung flux at GeV-energies, which could
account for the discrepancy between the measured GeV $\gamma$-ray
flux and predictions within the SSC-framework
\citep{1996ApJ...457..253D}. If this is true, one should observe,
in good approximation, a power-law spectrum between 100\,GeV and
10\,TeV with a spectral index 2.5-2.7. Another mechanism to be
mentioned is IC-scattering of relativistic electrons in the
unshocked pulsar wind. If the target photons are emitted by the
pulsar, a pulsed component could extend to $\gamma$-ray energies
of several 100\,GeV \citep{2000MNRAS.313..504B}. An independent
measurement in the intervening region between 60 and 400\,GeV
would constrain further the parameters of various models. The
MAGIC imaging atmospheric Cerenkov telescope (IACT) has a low-energy
trigger threshold ($\sim$50 GeV) and is currently the only
experiment capable of exploring this energy regime.

The spatially resolved morphology of the nebula is of a complex
nature. Its size in optical bandwidths is about 4$^{'}$ $\times$
6$^{'}$.  Due to synchrotron losses, the high-energy electrons
will have shorter cooling times and only the lower energy
electrons will reach out farther into the nebula. Thus, the
effective source size is expected to shrink with increasing energy
of the radiation. The radio emission is expected to extend up to
and beyond the filaments optically visible, whereas X-ray and
multi-TeV $\gamma$-rays should be produced in the vicinity of the
shock. On the other hand, the expected source size would increase
if the presence of an ionic component is established. In a special
study the HEGRA collaboration concluded that the rms size of the
VHE $\gamma$-ray emission region is $<1.5$\arcmin~for energies
above $1\,$TeV \citep{2000A&A...361.1073A}. A similar study at
energies below 1\,TeV has not been performed up to now.

The Crab pulsar is a source of pulsed radiation and has been
detected up to GeV energies. The Crab pulsar has a period of
$33\,$ms and was first discovered in radio
\citep{1968Sci...162.1481S} and shortly afterwards as the first
pulsar in the optical domain \citep{1969Natur.221..525C}. Since
then, pulsed emission from the Crab pulsar has been detected at
all accessible energies up to $\gamma$--rays \citep[for a
compilation of the broad band emission see][]{1999ApJ...516..297T}.
EGRET detected the Crab pulsar in $\gamma$-rays up to energies of
$10\,$GeV with a hint of a cutoff in the energy spectrum in the
highest energy bin at $6\,$GeV \citep{1993ApJ...409..697N}.
Despite various efforts, observations from the ground at higher
energies have, so far, failed to detect pulsed emission
\citep[e.g.][]{1999ICRC....3..460M,1999A&A...346..913A,2000ApJ...531..942L,2002ApJ...566..343D,2004ApJ...614..897A}.
Some experiments reported episodic pulsed emission
\citep{1982Natur.296..833G,1986Natur.319..127B,1992A&A...258..412A}
and persistent pulsed emission over a 1 yr period
\citep{1984ApJ...286L..35D}, but these observations have not been
confirmed by other experiments.

The high-energy emission from the pulsar is assumed to be due to
curvature and synchrotron radiation from relativistic charged
particles that are forced to move along magnetic field lines
inside the magnetosphere of the pulsar. The question of where the
particles are being accelerated is the subject of ongoing
theoretical activities. In the two most popular models, the
production of electrons and positrons and their acceleration take
place either above the polar cap of the neutron star
\citep[e.g.][]{1978ApJ...225..226H,1982ApJ...252..337D} or in
outer gaps in between the null surface and the light cylinder of
the magnetosphere
\citep[e.g.][]{1986ApJ...300..522C,1986ApJ...300..500C,1992ApJ...400..629C}.
We should not omit the slot-gap model, which places the
acceleration zone at the outer rim of the polar cap
\citep{1983ApJ...266..215A,2003ApJ...588..430M}. These models
differ in the predicted shape and cutoff of the energy spectrum at
the highest energies. A measurement of the turnover in the
spectrum would shed light on the possible sites of the particle
acceleration and can constrain models.

In this paper we report about the observation of the Crab Nebula
and pulsar with the MAGIC telescope between 2005 October and
December. After describing the MAGIC telescope and the
performed observations in \S\S \ref{magic} and \ref{selectio}, we present the
analysis chain in \S \ref{chain}. In \S \ref{dc} we
present our results on the steady emission comprising the
differential flux between 60\,GeV and 9\,TeV, a study of the
morphology of the emission region, and the measured integral flux
above $150\,$GeV. In \S \ref{ac} we present the results of
our search for pulsed emission, and we close with a discussion of our
results in \S \ref{discussion}.

\section{The MAGIC Telescope}\label{magic}

The MAGIC (Major Atmospheric Gamma Imaging Cerenkov) telescope
\citep[see][]{2004NewAR..48..339L} is located on the Canary
Island, La Palma (2200 m asl, $28.45^\circ$ north, $17.54^\circ$ west).
MAGIC is currently the largest single-dish IACT. It has a 17\,m diameter tessellated
reflector consisting of 964 $(0.5\times0.5)\,\mbox{m}^2$
diamond-milled aluminium mirrors, which are grouped onto support
panels in units of four. Depending on the elevation of the
telescope, the position of every panel is adjusted by
computer-controlled actuators of the so-called Active Mirror
Control, thus providing optimal focusing. About 80\% of the light
from a point source is focussed within a radius of $0.05^{\circ}$
in the focal plane. The MAGIC telescope is focused to a distance
of 10\,km---the most likely location of the shower maximum for
50\,GeV $\gamma$-ray air showers at small zenith angles.

The faint Cerenkov light flashes produced in air showers are
recorded by a camera comprising 577 photomultiplier tubes (PMTs).
The inner part of the camera (radius $\sim1.1^\circ$) is equipped
with 397 PMTs (type ET 9116A; 1\arcsec\,$\varnothing$ from Electron
Tubes Ltd [ET]) with a diameter of $0.1^\circ$ each . The outer
part of the camera is equipped with 180 PMTs (type ET 9117A;
1.5\arcsec\,$\varnothing$) of a diameter of $0.2^\circ$. The
central PMT is modified for optical pulsar studies
\citep{2005ICRC....5..367L}. Hollow hexagonal non-imaging light
concentrators (often called light catchers) are placed in front of
all photomultipliers to compensate for the dead space between them
and in order to shrink the observation solid angle to the
reflector. The entrance window of the PMTs is coated with a
diffuse lacquer doped with a wavelength shifter \citep[WLS;][]{2003NIMPA.504..109P}. The combination of the
hemispherically shaped PMT, the light catcher, the diffuse coating
and the WLS results in a 15\%-25\% higher quantum efficiency
compared to flat window PMTs. For protection purposes (humidity,
dust) a thin entrance window made of Plexiglas (type UG-218, with
a UV cutoff around $290\,$nm) is placed in front of the camera.

The PMTs have 6 dynodes and operate at a gain of roughly 30,000 to
slow down aging and damage from high currents by light during
observations close to the Galactic plane and during moonshine.
After amplification, the fast analogue signals are converted to
optical signals and transported by 160\,m long optical fibers to
the counting house. There the optical signal is  converted back
and split. Part of the signal is routed to the trigger. The
current configuration of the MAGIC camera has a trigger region of
2.0$^\circ$ in diameter \citep{2005ICRC....5..359C}. This provides
a $\gamma$-ray trigger collection area of the order of
$10^5\,$m$^2$ (at 200\,GeV for a source close to zenith).
Presently, the trigger energy range spans from 50-60 GeV (peak in
the differential trigger rates at small zenith angles for a
$\gamma$-ray source with a spectral slope of -2.6) up to tens of
TeV. An event is triggered if the signals in each of 4
neighbouring pixels exceed a threshold of $\sim7\,$photoelectrons (phe)
within a coincidence time window of $6\,$ns.

Before being digitized by an 8-bit, 300 MSamples/s FADC system,
each signal is stretched to an FWHM of about 6\,ns. The FADC
continuously writes the digitized amplitude information into a
ring buffer. In case of a trigger the digitization stops and the
corresponding part of the ring buffer is written onto a disk. The
dead time introduced by the readout is 25 $\mu$s. In order to
expand the dynamic range  to $\sim1000$, the signal of every PMT
is split into two branches, differing by a factor of 10 in gain.
The higher gain branch is read out for a $50\,$ns time interval.
When the signal amplitude exceeds a preset threshold, the delayed
 lower gain is routed to the same FADC channel and recorded in the following
$50\,$ns. Otherwise, the signal from the high gain branch continues
to be recorded and is
 used to determine the pedestal offset of each PMT channel.

The accuracy in reconstructing the direction of incoming
$\gamma$-rays on an event-by-event basis, hereafter $\gamma$-ray
point spread function or $\gamma$-PSF, is about 0.1$^\circ$,
depending on the energy. With the information provided by a
starguider camera, mispointing is corrected to an absolute
precision of about 1\arcmin. A $\gamma$-ray source with an
absolute intensity of $\sim2$\% of the Crab Nebula and similar
spectrum can be detected with MAGIC within 50 hr at energies
$>200\,$GeV on a significance level of $5\,\sigma$.

\section{Observations and Data Selection}\label{selectio}

Observations of the Crab Nebula with MAGIC are conducted on a
regular basis, as a means to monitor the performance of the
telescope. In this report we restrict ourselves to the analysis of
data obtained in the first observation cycle of the MAGIC
telescope between 2005 October and December. The observations were
performed in the so-called ON/OFF mode. The telescope was pointed
towards the Crab pulsar (ON) for about 16 hr. An OFF source
position, a sky region where no $\gamma$-ray source is known,  was
observed in the same range of zenith angles as the ON source. For
the background estimation we used OFF data collected for over 19
hr.

One of the main objectives of this analysis was to explore the
lower energy range of accessible $\gamma$-ray energies. The energy
threshold of IACTs  depends strongly on the zenith angle of
observation. Restriction to events with low zenith angles provides
the lowest possible energy threshold. Therefore, we select events
with zenith angles $\lesssim20^{\circ}$. For any given night the
data  affected by technical problems or fluctuations in the data
rate in excess of $10\,$\% were rejected. The atmospheric
conditions were judged from the nightly averaged and publicly
available atmospheric extinction coefficients from the nearby
Carlsberg Meridian telescope\footnote{See http://www.ast.cam.ac.uk/~dwe/SRF/camc\_extinction.html}. Within the selected
nights, the atmospheric light transmission changed by less than
5\%. The nights with data that survived all the selection
criteria, together with the corresponding observation times,
trigger rates, and zenith angle range, are listed in Table
\ref{datasummary}. The selected data sample comprises 14 nights
amounting to a total ON observation time of 955 minutes ($\sim16$
hr).
\begin{table}[tb]
\def\arraystretch{1.}
    \centering
        \caption{\label{datasummary}Data Selected for Analysis
        }
 \begin{tabular}{rrrr}\hline\hline
        \multicolumn{1}{c}{\rule[0mm]{0mm}{ 3mm}Date} & \multicolumn{1}{c}{Rates} & \multicolumn{1}{c}{On Time} &
        \multicolumn{1}{c}{Zd--Range}\\
         \multicolumn{1}{c}{\rule[-2mm]{0mm}{ 0mm}(MJD)}           & \multicolumn{1}{c}{(Hz)} & \multicolumn{1}{c}{(minutes)} & \multicolumn{1}{c}{(deg)} \\
        \tableline
        53,648...........................  & 130 &73 & $7-23$  \\
        53,655...........................  & 115 & 100 & $7-19$\\
        53,671...........................  & 122 & 105& $7-23$\\
        53,672........................... & 105 & 61 & $8-20$  \\
        53,679...........................  & 115 & 51 & $7-20$\\
        53,684...........................  & 95 & 50 & $7-20$  \\
        53,707...........................  & 98 & 53 & $7-11$  \\
        53,709........................... & 105 & 48 & $7-10$ \\
        53,711........................... & 105 &48 & $7-10$  \\
        53,713........................... & 108 & 50 & $7-14$  \\
        53,727........................... & 92 &44 & $7-13$ \\
        53,729........................... & 97 &61 & $7-13$ \\
        53,731........................... & 100 &107& $7-14$ \\
        53,735........................... & 92 & 104& $7-15$ \\\tableline
       \end{tabular}
        \tablecomments{The second column lists the average event rate after tail cuts and a cut in SIZE$>100\,$phe (for the
        definition of SIZE see text).}
    \end{table}

\section{Data Analysis}\label{chain}

The data analysis was carried out using the standard MAGIC
analysis and reconstruction software MARS \citep{2003ICRCMARS}.
After removing faulty and unstable camera channels, which amount
to 3\%--5\% of the total number of PMTs, the signal amplitudes,
extracted with the digital filtering method
\citep{magicextraction}, were converted to phe by
using the F-factor method~\citep{ffactororig}. Using calibration
events recorded interlaced to normal events
~\citep{magiccalibration}, the conversion factors were updated
every 10~s. There is a 10\% systematic uncertainty in the
calibration that directly propagates to the uncertainty of the
event energy scale (point 7 in Table \ref{syserrors} below). Time
offsets between pixels are corrected with a precision of better
than $1\,$ns.

\begin{figure}[b]
        \centering
                    \includegraphics*[angle=0,width=0.9\columnwidth]{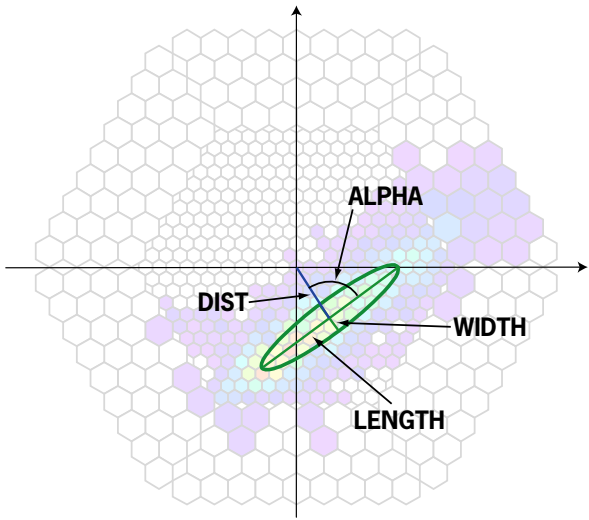}
                        \caption{Parameterization of a shower image with Hillas-parameters.}
                        \label{Hillas_par_fig}
     \end{figure}

After calibrating the data, pixels with faulty reconstructed
signal amplitudes and times were rejected and the corresponding
amplitude and time information was interpolated from the signals
and times of neighbouring pixels. Before image parametrization, a
tail-cut  cleaning of the image  was performed, requiring signals
higher than a pre-defined absolute amplitude level and time
coincidences (3.3\,ns) with neighbouring channels. The time
coincidence effectively suppresses pixels containing only a signal
from the night sky. For most of the analysis the minimum required
pixel content is 6\,phe for so-called core pixels and 4\,phe for
boundary pixels. For the morphology studies the minimum pixel
contents were raised to 10\,phe (core) and 8\,phe (boundary),
respectively, which improves the angular resolution, albeit at the
expense of an increased analysis threshold.

\begin{figure}[t]
        \centering
                    \includegraphics*[width=1.0\columnwidth]{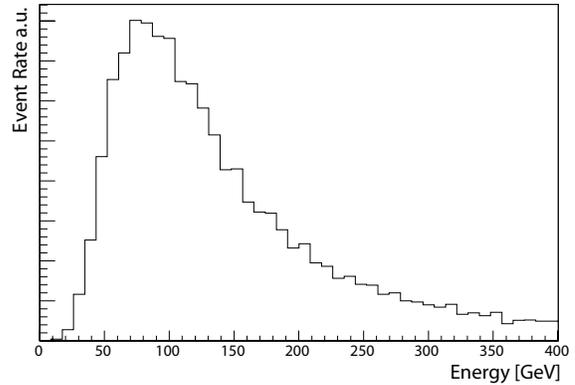}
                        \caption{Energy distribution of MC-$\gamma$-ray events with SIZE$>100$ phe for a simulated $\gamma$-ray source with spectral slope -2.6.}
                        \label{energy_distr}
     \end{figure}

Every cleaned event was parameterized by a principal
component analysis, commonly referred to as Hillas
parameterization~\citep{1985ICRC....3..445H}. The parametrization
was later used to separate between $\gamma$-ray event candidates
and background event candidates. The Hillas-parameters
DIST, LENGTH, WIDTH and also
ALPHA are illustrated for a recorded shower image in
Figure \ref{Hillas_par_fig}. Another useful parameter is the
SIZE of a shower image, the intensity of the image after
image cleaning in units of recorded photoelectrons. Note that
SIZE depends on the applied tail cuts. SIZE is a
good estimate of the primary particle energy, provided that the shower
impact distance to the telescope principal axis is below
$\sim120\,$m. An event pre-selection was performed by discarding
event candidates affected by noise and pick-up (e.g.~car flashes)
and event candidates with a low number of pixels (after tail-cuts
typically a minimum number of 5 core pixels were requested). In
addition, an image-SIZE of at least $\sim100\,$phe was
requested. Figure \ref{energy_distr} shows the energy distribution
of simulated $\gamma$-ray events with SIZE $>100\,$phe.
The distribution peaks at an energy of $75\,$GeV. The simulated
$\gamma$-ray source  has a powerlaw spectrum with an index of
-2.6.

\begin{figure*}[t!]
        \centering
                    \includegraphics*[width=1\columnwidth]{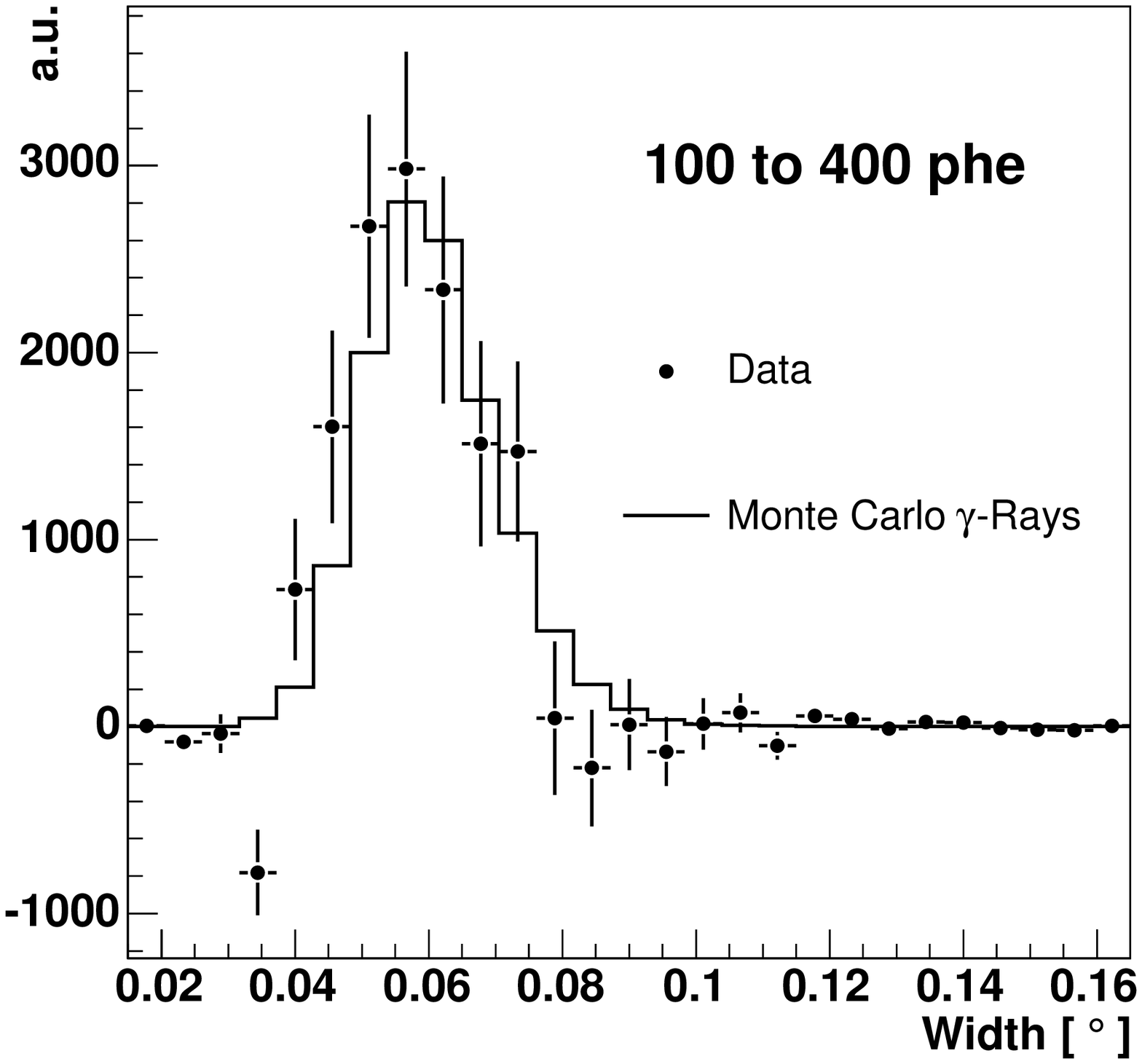}
                    \includegraphics*[width=1\columnwidth]{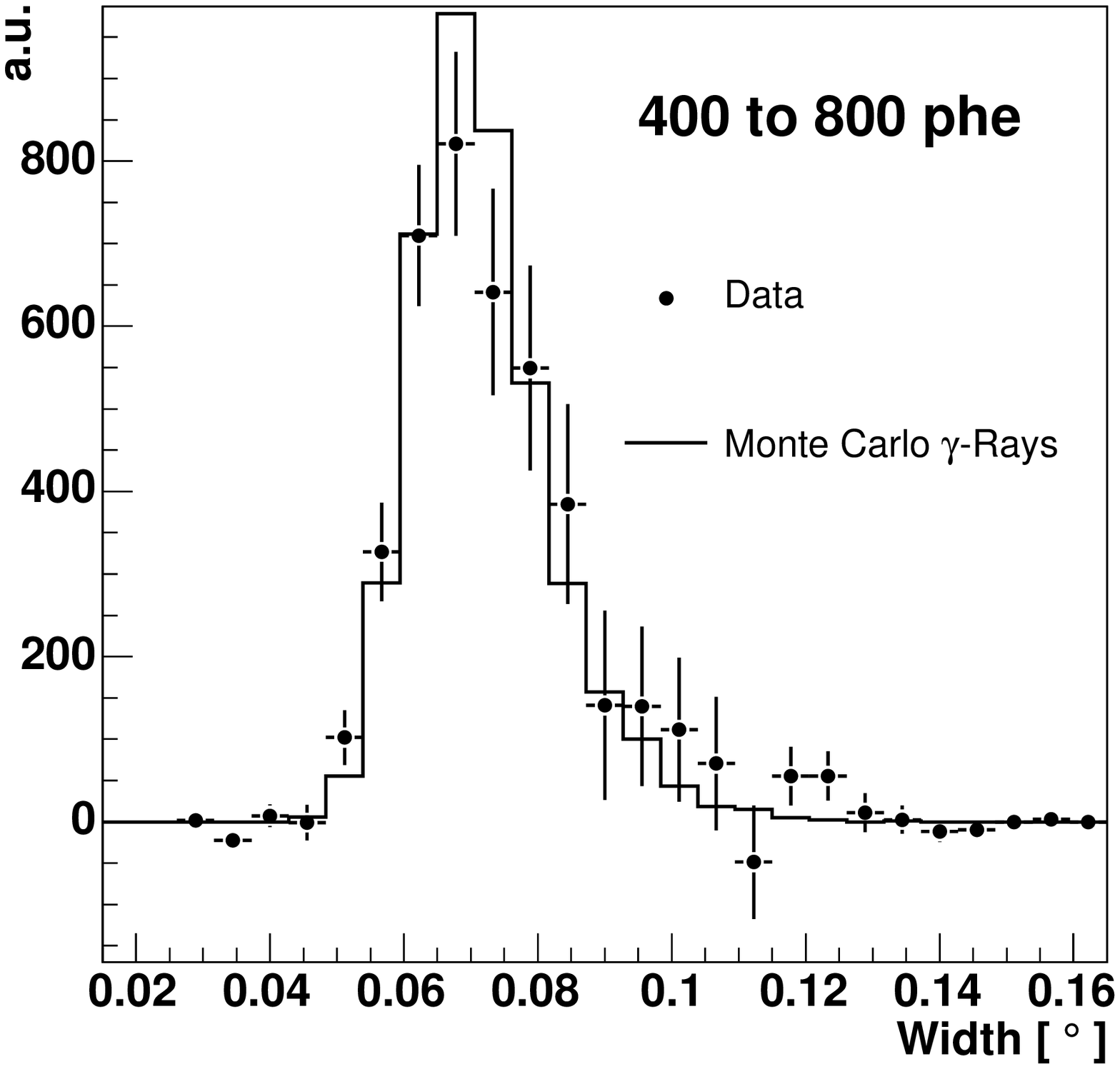}\\
                    \includegraphics*[width=1\columnwidth]{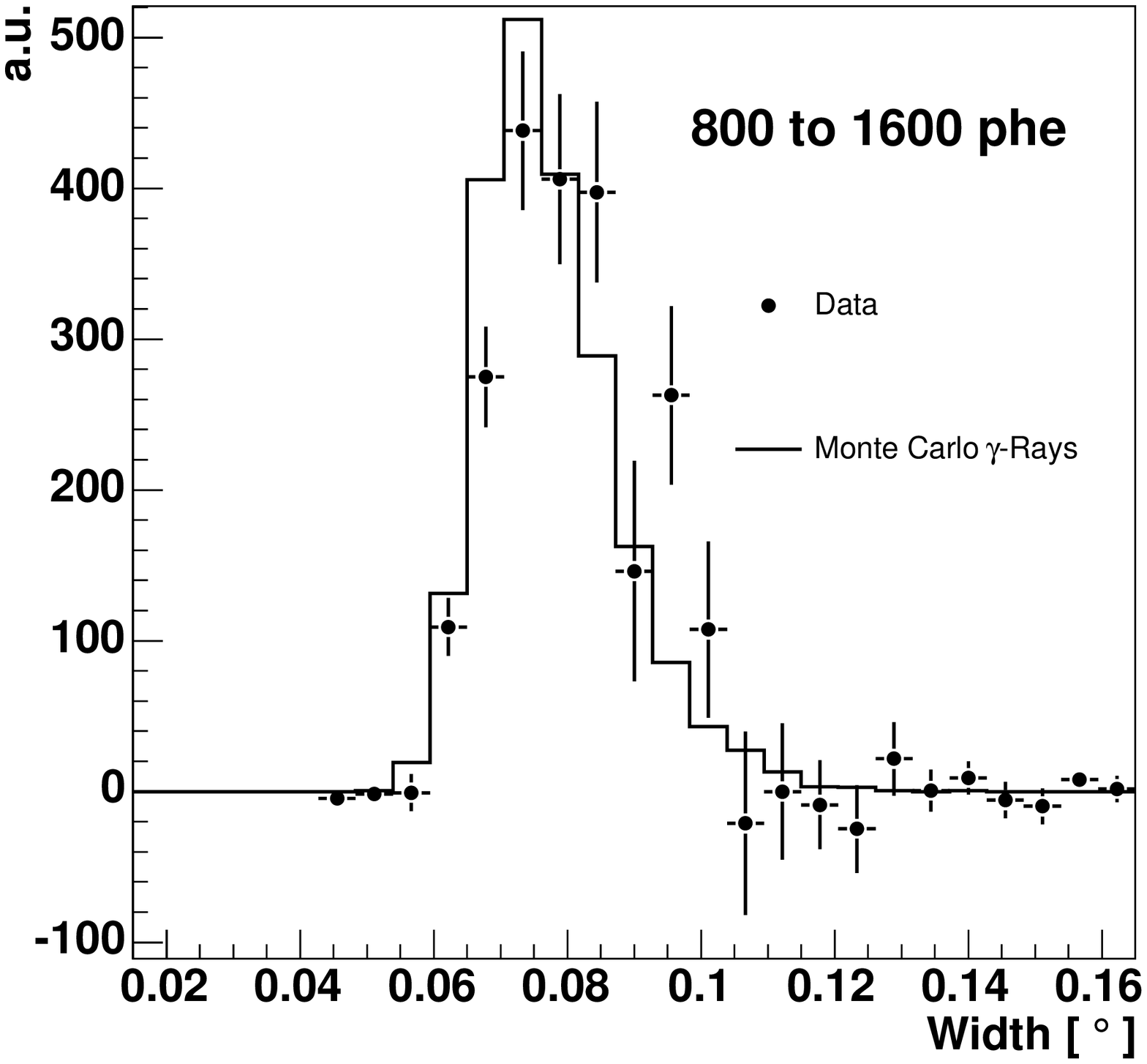}
                    \includegraphics*[width=1\columnwidth]{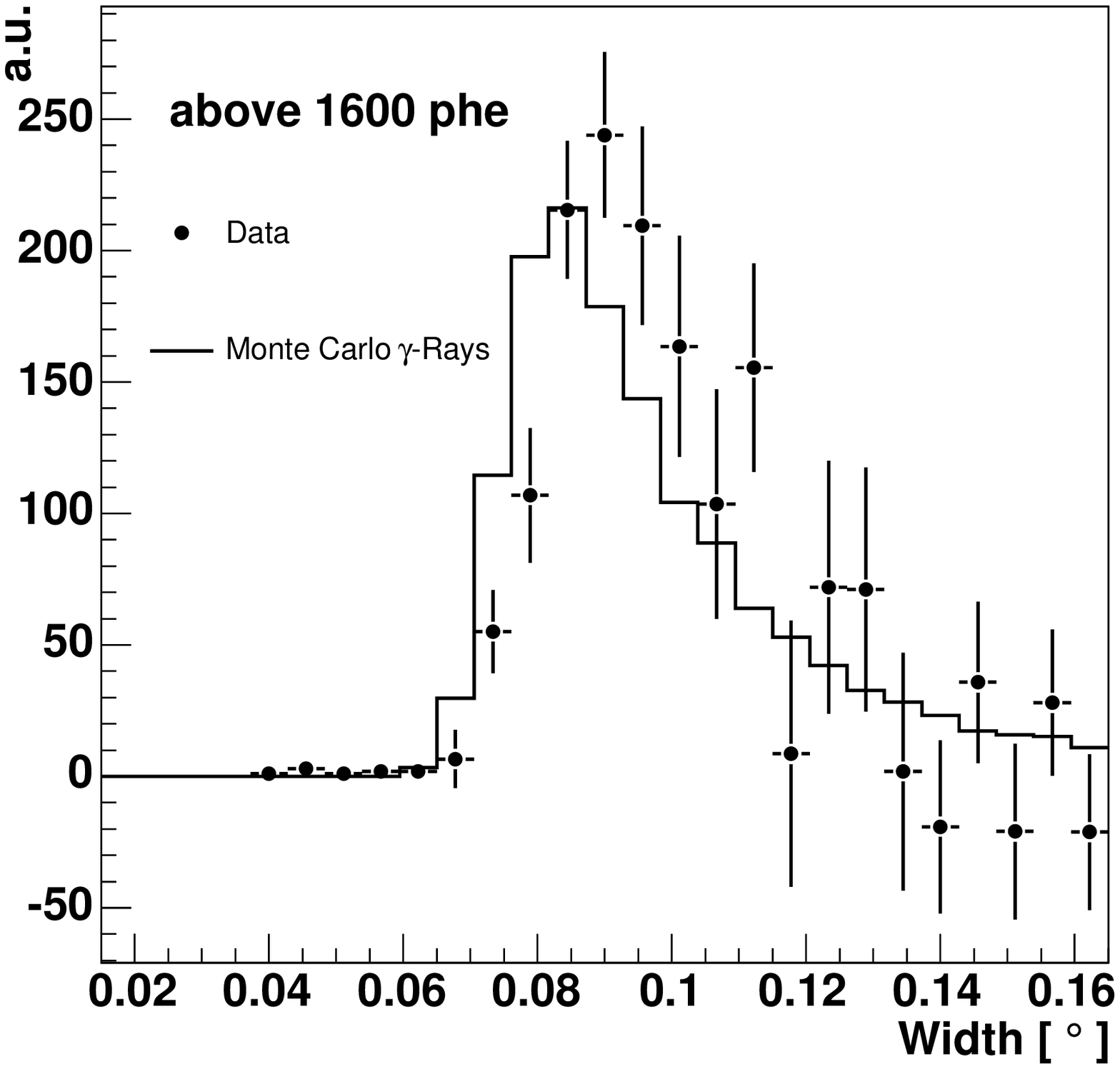}
                        \caption{Distribution of the image parameter WIDTH for
                        $\gamma$-ray events extracted from data
                        and for MC simulated $\gamma$-rays in four consecutive bins of
                        SIZE covering the full analyzed energy range.}
                        \label{MC_data_comparison}
     \end{figure*}

A Monte Carlo (MC) simulation properly describing data
is a necessary requisite for a ground-based $\gamma$-ray
experiment. In Figure \ref{MC_data_comparison} the image parameter
WIDTH is shown for simulated $\gamma$-rays and
$\gamma$-rays extracted from data. The four panels are for
consecutive bins in SIZE covering the entire range of
analyzed $\gamma$-ray energies. For an unbiased comparison, loose
cuts have been applied in the $\gamma$/hadron-separation
(explained in the next section). The $\gamma$-ray excess  was
obtained by subtracting the scaled distribution of the OFF-data
sample from the distribution of the ON-data sample. The scaling
factor was found by normalizing the
$|\mbox{ALPHA}|$-distributions of both samples between
$|\mbox{ALPHA}|$=30$^{\circ}$ and 70$^{\circ}$. The comparison
was done by selecting events with small $|\mbox{ALPHA}|$
(typically less than $10^{\circ}$). A small $|\mbox{ALPHA}|$ is
expected for $\gamma$-rays from the Crab Nebula, as
explained later. The agreement of the MC simulated
distributions and the distributions extracted from data is
acceptable in all four SIZE-bins. However, it is evident
from the figures that the agreement worsens at large
SIZE. In the analysis, a possible bias introduced by this
behavior is avoided by applying rather loose cuts at energies
above several TeV.

\subsection{Gamma Hadron Separation}

Only a small fraction between $10^{-3}$ and $10^{-4}$ of the
recorded data are $\gamma$-ray showers. The major fraction of the
recorded events are cosmic-rays of hadronic origin. This unwanted
background has to be suppressed offline. For this purpose  we
applied a multivariate method, the random
forest~\citep{2001breiman,2004NIMPA.516..511B,magicrandomforest},
which uses the image-parameters to compute the HADRONNESS
of an event. The HADRONNESS of an event quantifies the
probability of an event to be ``$\gamma$-like'' or
``hadron-like.'' The random forest is trained with MC
simulated $\gamma$-ray events and either with simulated hadronic
cosmic-ray events or, as in the present case, with background
events recorded by MAGIC. In this study, we used for the training
of the random forest the image parameters SIZE,
DIST, WIDTH and LENGTH, as well as the
third moment along the major axis of a shower image, and a
parameter describing the CONCENTRATION of a shower image.

\begin{figure*}[t]
   \centering
                    \includegraphics*[width=1.6\columnwidth]{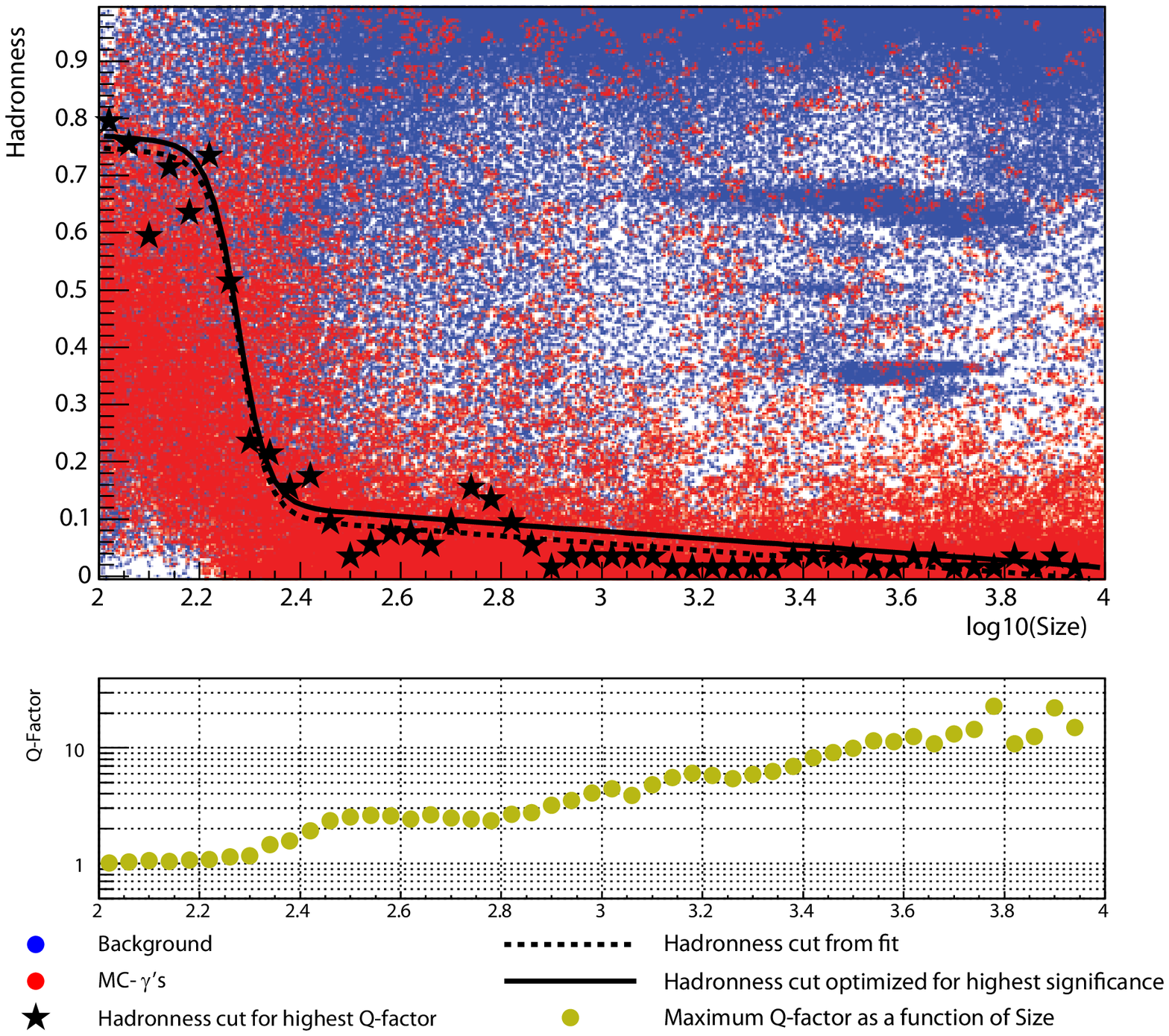}
                        \caption{\emph{Top}: Distribution of the parameter
                        HADRONNESS vs.~SIZE. The background
                        events are marked blue
                        and the MC $\gamma$-ray events red. Note that below a
                        SIZE of 300\,phe most of the background events are
                        hidden behind the
                        simulated $\gamma$-ray events.
                        For bins in SIZE the
                        HADRONNESS cuts  that yield
                        the highest quality factors (\emph{black stars}) are found.
                        The dashed line is a fit to the
                         HADRONNESS cut
                        values. The solid line is the fit shifted by a constant value and results in the
                        $\gamma$-ray signal from the Crab Nebula with the highest significance.  \emph{Bottom}: Quality
                        factor for the HADRONNESS cut (\emph{stars}) in the top panel.}
                        \label{hadr_cut}
   \end{figure*}

Figure \ref{hadr_cut} shows the HADRONNESS for MC simulated $\gamma$-ray showers (\emph{red}) and for recorded
background (\emph{blue}) as a function of SIZE. A clear
separation between both populations is visible for SIZE
$\gtrsim300\,$phe, corresponding to $\gamma$-ray energies
$\gtrsim150\,$GeV. Below 300\,phe both populations start to
overlap until, at $\sim200\,$phe ($\sim100\,$GeV), no more
separation is possible. The bottom panel in the figure shows the
maximum quality factor \emph{Q}
$(\epsilon_\gamma/\sqrt{\epsilon_\mathrm{B}})$ for each
SIZE interval obtained for an optimized
HADRONNESS cut. Parameter $\epsilon_\gamma$ is the fraction of
retained $\gamma$-ray events and $\epsilon_\mathrm{B}$ the
fraction of retained background events. The corresponding
HADRONNESS cut is shown by the stars in the top panel
of the figure.

\begin{figure*}[t]
        \centering
         \includegraphics*[bb = 140 53 574 716, angle=-90,
                    width=0.9\columnwidth]{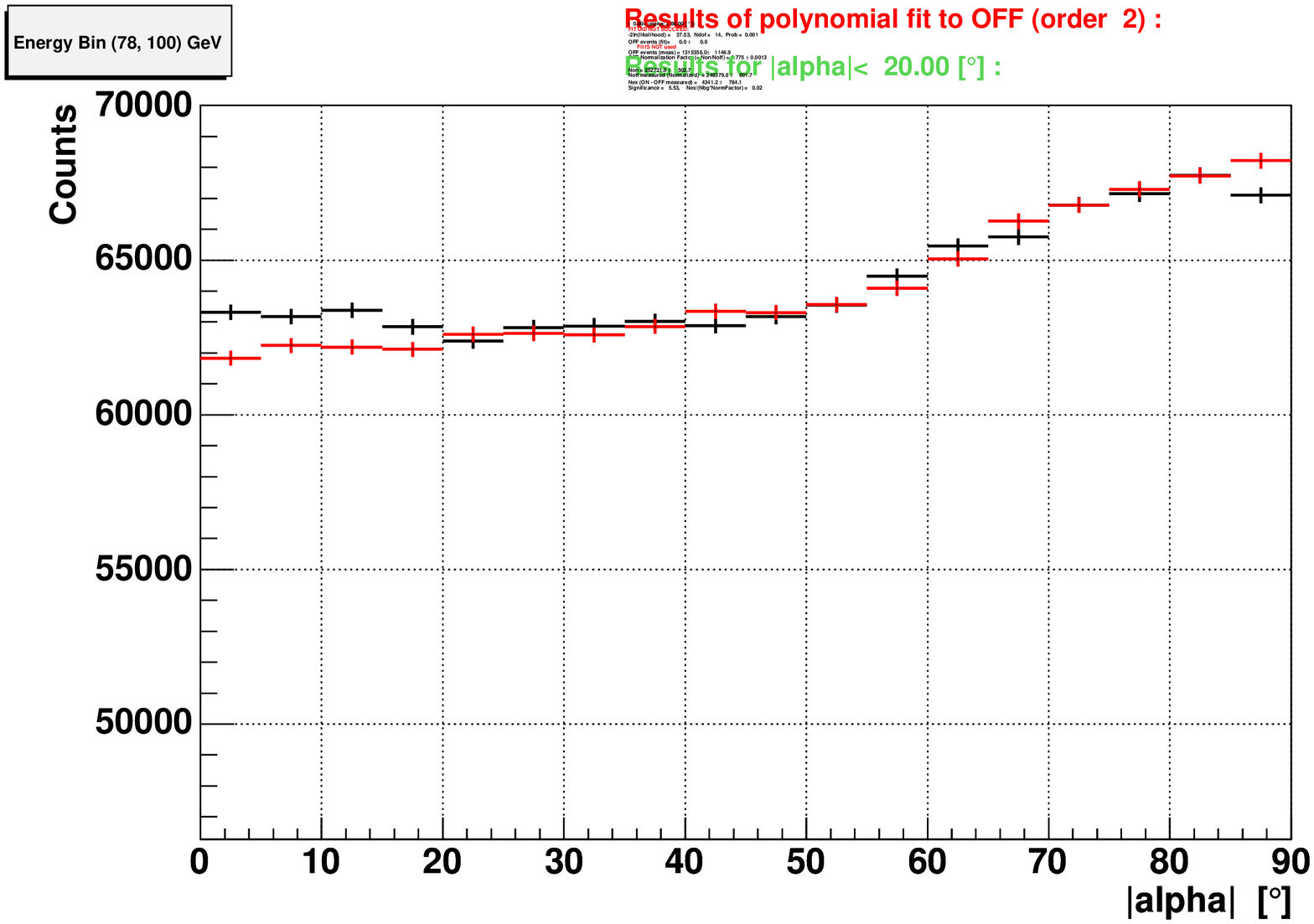}
         \includegraphics*[bb = 140 38 551 740, angle=-90,width=1.05\columnwidth]{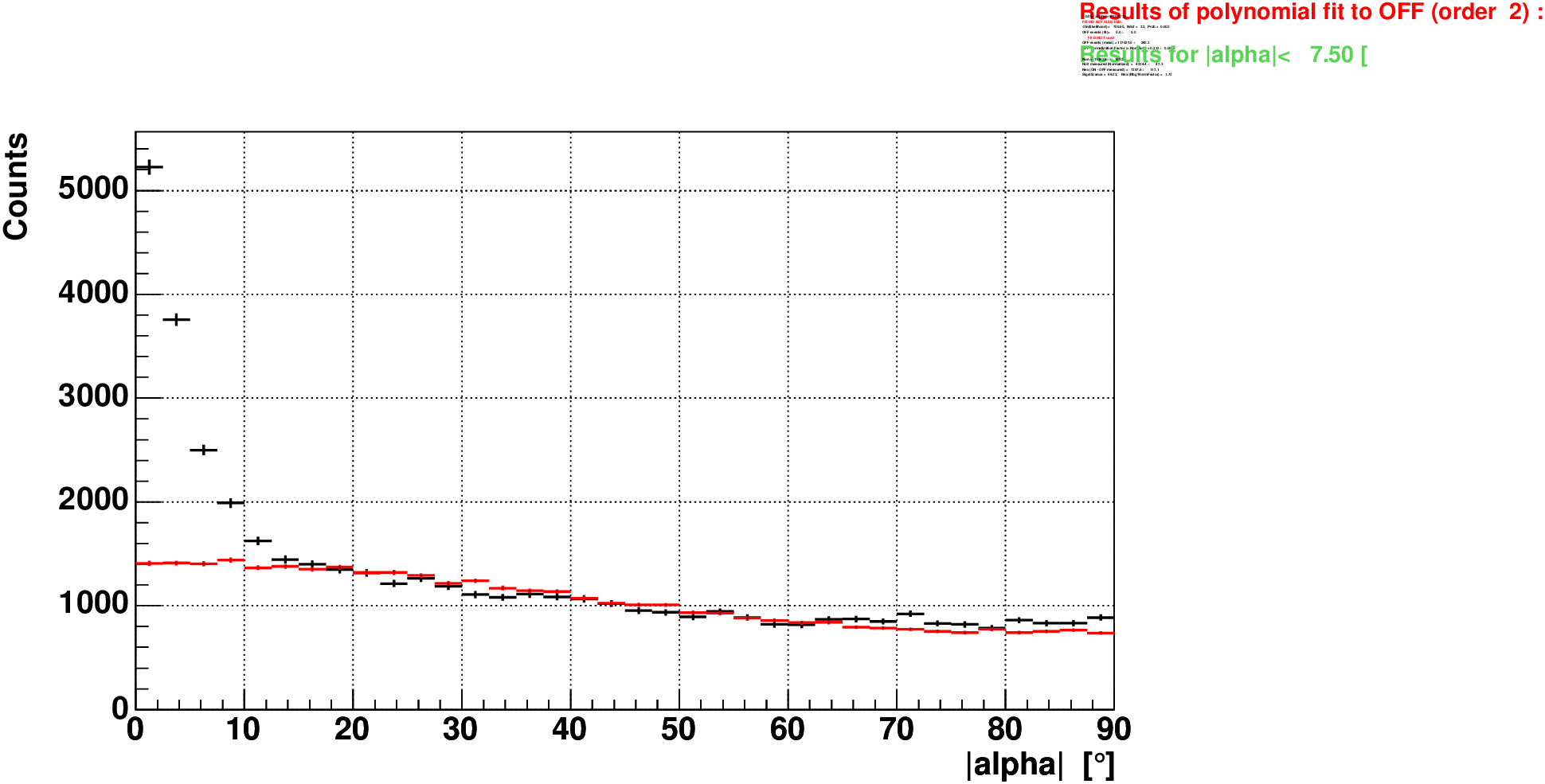}
                        \caption{Distribution of the image parameter $|\mbox{ALPHA}|$
                         for the bin of reconstructed energy $80-100\,$GeV
                        in the left panel and for events with energies $>200\,$GeV
                         in the right panel. (\emph{black}: ON-data; \emph{red}: OFF-data)}
                        \label{crab:alphabelow100GeV}
  \end{figure*}

The random forest method is also used to estimate the energy of
each event~\citep{magicrandomforest}. An energy resolution of
$\sim25\%$ is achieved for events with energies $>200\,$GeV. The
energy resolution  reduces to $\sim40$\% around $70\,$GeV.

\subsection{Signal Extraction}

After $\gamma$/hadron-separation and  energy estimation, the
$\gamma$-ray signal is extracted from an
$|\mbox{ALPHA}|$-distribution. ALPHA is the angle
between the major axis of the recorded shower and the vector
connecting its center of gravity (CoG) with the source position in
the camera plane (cf. Figure \ref{Hillas_par_fig}). Shower images
of $\gamma$-rays from the source point with their major axes
toward the source position in the camera and appear as an excess
at small values in the $|\mbox{ALPHA}|$-distribution. Figure
\ref{crab:alphabelow100GeV} shows two
$|\mbox{ALPHA}|$-distributions from the data (\emph{black:} ON-data;
\emph{red}: OFF-data). The left panel in Figure
\ref{crab:alphabelow100GeV} shows the
$|\mbox{ALPHA}|$-distribution of events with estimated energies
between 80 and $100\,$GeV; the right panel shows the
distribution of events with reconstructed energies above
$200\,$GeV. In both cases an excess of $\gamma$-ray events  is
clearly visible. However, the significance of the $\gamma$-ray
signal at lower energies  is considerably reduced compared to
higher energies  because of the degradation of the background
suppression towards lower energy (cf.~Figure \ref{hadr_cut}).

\begin{figure}[t]
        \centering
                    \includegraphics*[width=\columnwidth]{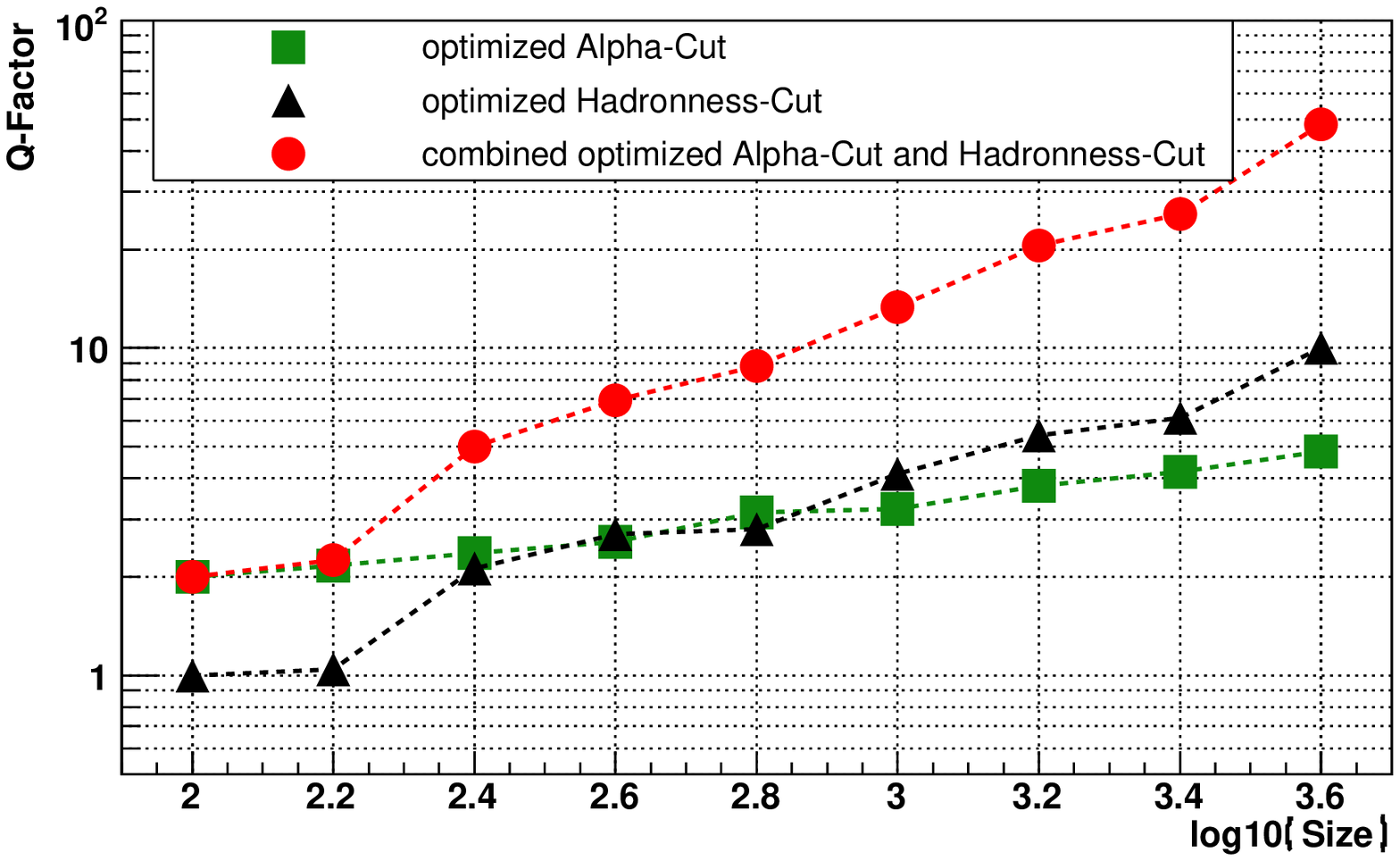}
                        \caption{Size dependence of the \emph{Q}-factor for optimized $|\mbox{ALPHA}|$ cut and
                         HADRONNESS cut, and the \emph{Q}-factor for the combination of $|\mbox{ALPHA}|$ cut and HADRONNESS cut.}
                        \label{comb_q_factor}
  \end{figure}

At the lowest energies ALPHA is currently the only means
by which it is possible to separate $\gamma$-rays and background
events. This is illustrated in Figure \ref{comb_q_factor}, which
shows the quality factor as a function of SIZE separate
for an optimized ALPHA cut and optimized
HADRONNESS cut, as well as the combination of both cuts.
Below SIZE 250\,phe a cut in HADRONNESS does not
improve $\gamma$/hadron-separation and reduces only statistics. On
the other hand, with ALPHA a quality factor of 2 is
still possible.

\subsection{Sensitivity}

The highest integral sensitivity for $\gamma$-ray emission from
the Crab Nebula is obtained above $\sim250\,$GeV. The integral
sensitivities for energies above $250\,$GeV obtained on a daily
basis are listed in Table \ref{intesensi}. The sensitivities are
calculated in units of
$\sigma_\mathrm{LiMa}/(\mathrm{hours})^{1/2}$; the significance
$\sigma_\mathrm{LiMa}$ is calculated using equation 17 from
\cite{1983ApJ...272..317L}. In the last column of the table, the
sensitivity is expressed as the minimum flux, normalized to the
Crab flux that can be measured with 5 $\sigma$
significance\footnote{Note that in this case the significance is
calculated as\\ excess / (background)$^{1/2}$.} in a 50 hr
observation. The day-by-day sensitivities vary by about $10\%$,
indicating a stable telescope performance throughout the
observations.

The energy dependence of the sensitivity was studied by
calculating the integral sensitivity  for several analysis
thresholds. The SIZE-dependent HADRONNESS cut
(Figure \ref{hadr_cut}, \emph{black solid line}) was used for
$\gamma$/hadron separation and the ALPHA cut was
tightened with increasing energy based on studies from
MC-simulations.
%shown in Figure \ref{alpha_cut}.
The integral sensitivity for a 50 hr observation is 13\% Crab
at 75\,GeV and  improves continuously with increasing the analysis
threshold to about 2.2\% above $\sim250\,$GeV (see Table
\ref{int_sens_table}). Note that above $1.6\,$TeV the background
is estimated from only two events, which results in an uncertainty
of more than $70\%$ on the integral sensitivity.

\begin{table}[htb]
    \centering
    \caption{\label{intesensi}MAGIC Integral Sensitivities
        to $\gamma$-Ray Emission\\ from the Crab Nebula for $\gamma$-Ray Energies above
        250\,GeV}
       \begin{tabular}{ccc}\hline\hline
        \rule[0mm]{0mm}{ 3mm}{Date} &
        {Sensitivity in}& {Sensitivity in}\\
          \rule[-2mm]{0mm}{ 0mm}\mbox{}(MJD)     & [$\sigma_\mathrm{LiMa}/(\mbox{hours})^{1/2}$] & (\% Crab, 50 hr, 5$\sigma$)\\
        \hline
        53,648...................   & 18.1 & 2.4\\
        53,655...................  & 19.6 & 2.1\\
        53,671...................  & 19.9 & 2.1 \\
        53,672...................  & 19.1 & 2.2\\
        53,679...................   &19.4  & 2.2 \\
        53,684...................  & 21.1 & 1.8 \\
        53,707...................   &23.6 & 2.0 \\
        53,709...................   & 16.1 & 2.7 \\
        53,711...................   & 18.9 & 2.2 \\
        53,713...................   & 18.7 & 2.2 \\
        53,727...................   & 19.1 & 2.3 \\
        53,729...................   & 17.5 & 2.3 \\
        53,731...................   & 18.0 & 2.3 \\
        53,735...................   & 18.7 & 2.1 \\
                \hspace{5mm}Average..........   &  $19.1\pm0.5$ (1.7)&$2.21\pm0.05$ (0.2)\\ \tableline
       \end{tabular}
        \tablecomments{Average values of each column are given in the last row with the corresponding RMS value in brackets. Note that different definitions of significance are used in the two columns.}
    \end{table}

\begin{table*}[t]
    \centering
\caption{\label{int_sens_table} Integral Sensitivities of the MAGIC Telescope to the $\gamma$-Ray Emission\\ from the Crab Nebula for
       Several Analysis Thresholds
        }
       \begin{tabular}{ccrrrc}\hline\hline
        \rule[0mm]{0mm}{ 3mm}{Energy} & {$|\mbox{ALPHA}|$ Cut} &&&& {Sensitivity}\\
         \rule[-2mm]{0mm}{ 0mm}\mbox{(GeV)}   &  {(deg)} & {ON Events} & {OFF Events} & { Excess Events}&(\% Crab, 50 hrs, 5$\sigma$)\\
        \tableline
        $>75$......................... & $<10^\circ$ & 232505 & 221751 &10754 &13.0\\
        $>110$....................... & $<10^\circ$ & 58600 & 49702 &8898 &7.5\\
        $>200$....................... & $<10^\circ$ & 11399 & 4960 &6439 &3.3\\
        $>400$....................... & $<7.5^\circ$ & 2866 & 348 &2518 &2.2\\
        $>800$....................... & $<5^\circ$ & 613 & 20 &593 &2.3\\
        $>1600$..................... & $<5^\circ$ & 43 & 2 & 41&1.0\\\tableline
       \end{tabular}

    \end{table*}

\subsection{Systematic Uncertainties}

Apart from statistical errors, many results in cosmic-ray
experiments are affected by rather large systematic errors. One of
the main problems is the lack of test beams that allow calibration
of the entire instrument in combination with the showering process
in the atmosphere. The standard replacement for a test beam
calibration is the use of MC simulation based on many
linked processes either using physical models (example: the
simulation of electromagnetic showers) or calibrating them in
separate measurements (example: the spectral mirror reflectivity).
In some cases one has to make reasonable guesses (example:
photoelectron collection efficiency in the PMT front-end volume).
The calibration of individual effects suffers partly from
cross-correlations, which are not always well understood.
Currently, the best approach is to estimate the systematic
uncertainties (commonly called systematic errors) of the various
parameters separately and to combine them to a global systematic
error. Here, we follow the general practice of adding these
individual errors in quadrature although this will result in a
slight underestimate of the total systematic error. Table
\ref{syserrors} lists the dominant systematic error contributions
($\geq2\%$) for the spectral parameters (flux, slopes, cut-offs, etc.).
The systematic errors influence the spectrum in different ways.
Some (Nr 1-7, 10, 11) result in an uncertainty of the energy scale
and thus can enter as a large factor in the flux at a given energy
in case of steep spectra with slopes $<3$; others (Nr
8, 9, 12-13, 15-17) linearly influence the flux normalization for the
spectrum.

The most critical contributions to the
systematic error come from the uncertainties in the conversion of
photons to measurable photoelectrons (combined under item  7), the
so-called photon detection efficiency (PDE). The PDE is a
combination of many small effects such as the reflectivity
variation of the light catchers, tolerances in the light catcher
geometry, angular effects on the PMT surface, non-uniformity of
the diffuse lacquer coating, the QE-spread and cathode
non-uniformity of the PMTs, the photoelectron collection
efficiency in the PMT front-end volume, and gain variations of the
first dynode. Also some contribution of the signal transmission to
the DAQ is included.

Fortunately, the PDE can normally be measured with a light source
uniformly illuminating the camera with short blue or UV light
pulses. Obviously, the light pulser itself introduces some
systematic errors such as in the absolute light flux
determination, small deviations from uniformity in the
illumination, some (small) temperature drift, and amplitude jitter.
Also, the used method of determining the number of detected
photoelectrons, the above-mentioned so-called F-factor method,
introduces some uncertainty.

\begin{figure}[t]
        \centering
                    \includegraphics*[width=\columnwidth]{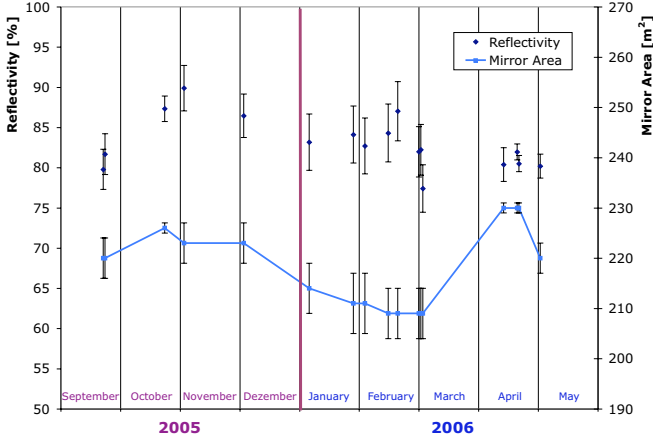}
                        \caption{Measurement of the reflectivity of the main
                         mirror; the effective mirror area is shown by the blue points \citep{garc07}. The measurements cover the
                         Crab observation period from 2005 October to December.}
                        \label{reflect}
  \end{figure}

Another rather big uncertainty is the effective reflectivity of
the mirrors defined as the light from a source at infinity being
focused onto the area of a pixel. Comparing the measured
brightness of a star and its image back-reflected by a high-quality diffuse reflector in the camera plane allows one to carry
out a routine reflectivity measurement (see Figure \ref{reflect}),
with an uncertainty of about 7\%.  A similarly large error
contribution was estimated for the event reconstruction. Again,
many small effects contribute to the reconstruction losses or to
the wrong assignment of events. In contrast to the procedure to
limit the uncertainty as in the example of the PDE, no simple
method to cross-check the error range of the reconstruction
efficiency is possible and a reasonable guess had to be made.

Effects that influence the slope of reconstructed $\gamma$-ray
energy spectra (class C effects in Table \ref{syserrors}) are
mostly dominating at the lowest and highest accessible energies.
The estimate of the systematic slope error is rather difficult. In
case of a power law or moderately curved power law we estimate an
uncertainty on the slope of 0.2. We note that measurements  by
current second-generation telescopes of the spectral slope of the
Crab Nebula agree better than 0.1 in the overlapping energy range.

In summary, we obtain a systematic energy scale error of 16\%, a
systematic error of 11\% on the flux normalization (without the
energy scale error), and a systematic slope error of $\pm0.2$
(which is a combination of error 13 and the other relevant class A
errors averaged over the energy).

\section{Analysis Results}\label{dc}

\subsection{Differential Energy Spectrum of the Crab Nebula}

\begin{figure}[b]
        \centering
         \includegraphics*[angle=0,
                    width=\columnwidth]{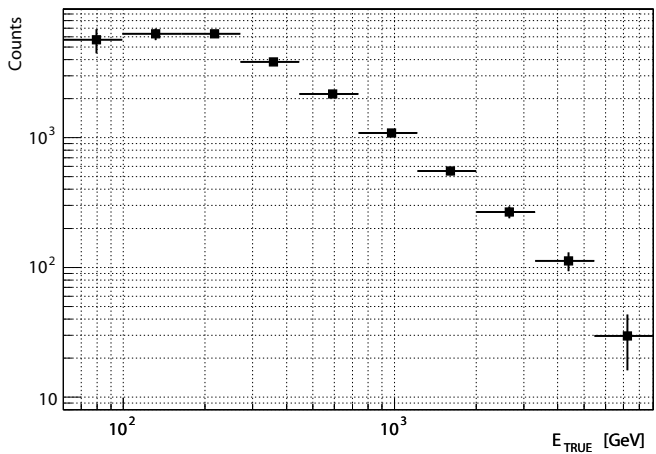}
                        \caption{Unfolded distribution of excess events from the Crab Nebula.
                         The integral rate of excess events is 0.4\,Hz.}
                        \label{crab:excess}
  \end{figure}

\begin{figure}[t]
        \centering
         \includegraphics*[angle=0,
                    width=\columnwidth]{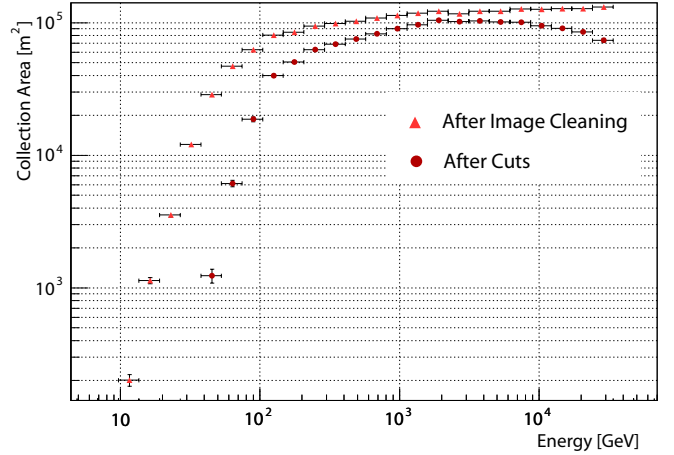}
                        \caption{Collection area after image cleaning and after cuts for low zenith angle observations ($<20^\circ$).}
                        \label{collarea}
  \end{figure}

\begin{figure}[t]
        \centering
         \includegraphics*[angle=0,
                    width=1.1\columnwidth]{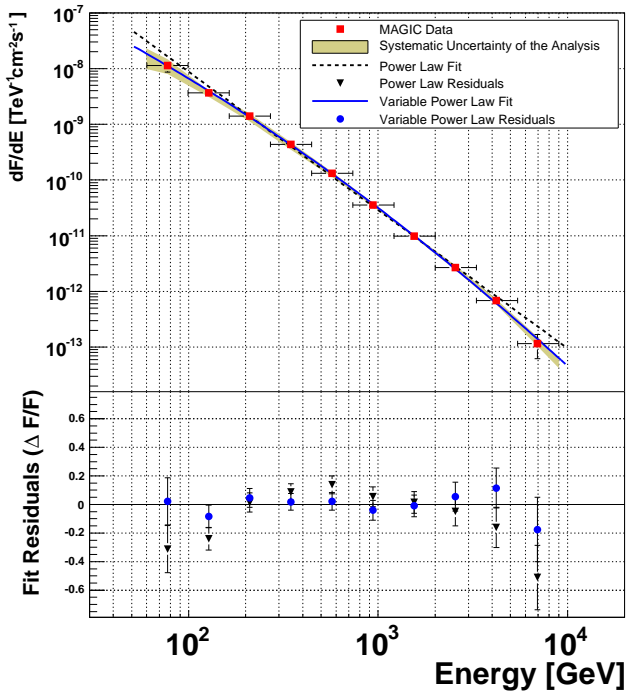}
                        \caption{Differential energy spectrum of the Crab Nebula. The spectrum was unfolded with the method
                    of \cite{Bertero88}.  The
                        results of a fit of the spectrum with a
                        power law and a broken power law are also shown.
                        The
                        bottom panel shows the relative residuals between
                        the fit and the data points. See text for further discussion.}
                        \label{crab:diffspectr}
  \end{figure}

By extracting the $\gamma$-ray signal in each bin of the
reconstructed energy $E_\mathrm{rec}$, a spectrum $N_i$ of
$\gamma$-rays in each $E_\mathrm{rec}$ bin $i$ can be constructed.
The reconstructed energy is subject to a bias. Before determining
a differential $\gamma$-ray flux in true energy $E_\mathrm{true}$
bins, the spectrum $N_i$ has, therefore, to be converted into a
spectrum $M_j$ of $\gamma$-rays in bins of $E_\mathrm{true}$. This
is done by applying an unfolding procedure with regularization
\citep{1991NIMPA.303..350A}. An essential input for the unfolding
procedure is the migration matrix, which describes the migration
of events from bin $i$ in $E_\mathrm{rec}$ into a bin $j$ of
$E_\mathrm{true}$. The migration matrix is determined from MC
simulated $\gamma$-ray showers. The unfolding is done
independently for different regularization schemes
\citep{Tikhonov79,Bertero88,1994NIMPA.340..400S,magicunfolding}.
Figure \ref{crab:excess} shows one distribution of excess events
from the Crab Nebula after unfolding by the method of
\cite{Bertero88}. The integral rate of excess events is 0.4\,Hz.
The differences between the unfolded points $M_j$ obtained with
the different regularization schemes are used to estimate a
systematic error due to the unfolding. Figure
\ref{crab:diffspectr} shows the differential $\gamma$-ray flux,
which was  obtained with the regularization scheme proposed by
\cite{Bertero88} and by normalizing the unfolded spectrum $M_j$,
to the effective collection area (Figure \ref{collarea}),  the
effective observation time, and the bin width of $E_\mathrm{true}$
(given by the horizontal bars at each flux point in the figure).
The average differential flux for each energy bin is presented in
Table \ref{diffpoints}.

The influence of different choices for tail-cuts,
HADRONNESS cuts, DIST cuts, and core-pixel cuts
on the measured flux is indicated in Figure \ref{crab:diffspectr}
by the shaded region and quoted as systematic uncertainty in Table
\ref{diffpoints}. Due to analysis uncertainties, the band broadens
at low energies, mostly because of limited $\gamma$/hadron
discrimination power. It broadens at the highest energies due to
low event statistics.

The energy spectrum is parameterized with both a power-law and a
curved power-law \emph{Ansatz}. The fit takes into account correlations
between the spectral points that are introduced by the unfolding
procedure. A correlated fit with a power-law
\begin{equation}
 \frac{\mathrm{d}F}{\mathrm{d}E}=f_0\,\left(E/300\,\mathrm{GeV}\right)^{\Gamma}\quad,
\end{equation}
provides a flux normalization $f_0$ of
$(5.7\pm0.2_{\mathrm{stat}})\times10^{-10}\,$cm$^{-2}$s$^{-1}$TeV$^{-1}$
and a spectral index $\Gamma$ of
$-2.48\pm0.03_{\mathrm{stat}}\pm0.2_{\mathrm{syst}}$. The $\chi^2$
of the fit is 24 for 8 degrees of freedom, which disfavors a pure
power-law description of the spectrum. The energy spectrum is
better described by a curved power-law \emph{Ansatz}
\begin{equation}
\frac{\mathrm{d}F}{\mathrm{d}E}=f_0\,
\left(E/300\,\mathrm{GeV}\right)^{\left[a+b\log_{10}\left(E/300\,\mathrm{GeV}\right)\right]}
\end{equation}
yielding a flux normalization $f_0$ of
$(6.0\pm0.2_{\mathrm{stat}})\times10^{-10}\,$cm$^{-2}$s$^{-1}$TeV$^{-1}$,
$a=-2.31\pm0.06_{\mathrm{stat}}$ and
$b=-0.26\pm0.07_{\mathrm{stat}}\pm0.2_{\mathrm{syst}}$. The
$\chi^2$ of the fit is 8 for 7 degrees of freedom.

\begin{figure}[b]
        \centering
                    \includegraphics*[width=1.1\columnwidth]{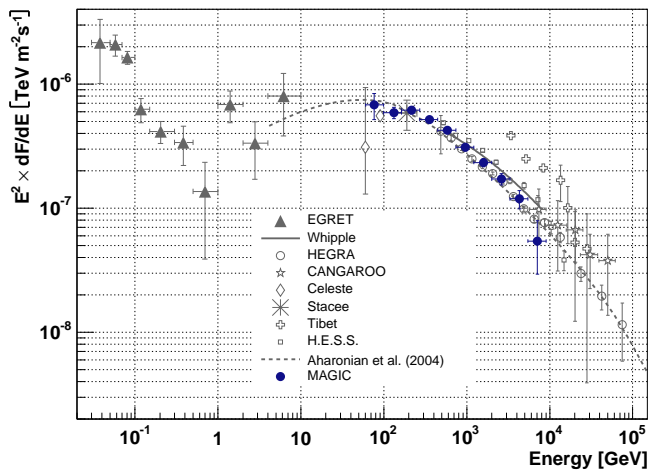}
                        \caption{SED of the $\gamma$-ray emission
                         of Crab Nebula.  The measurements shown below 10\,GeV are by EGRET \citep{1996ApJ...457..253D}.
                        In VHE $\gamma$-rays, measurements
                        are from ground-based experiments. Above 400\,GeV our measurement is in agreement with measurements by other
                        IACTs. The dashed line is a model
                        prediction by \cite{2004ApJ...614..897A}.
                        }
                        \label{Crab_sed_time}
  \end{figure}

Figure \ref{Crab_sed_time} shows the differential flux
measurements multiplied by the energy squared, i.e.~the spectral
energy distribution (SED). In the figure we compare our
measurement with those from other experiments. For energies above
400\,GeV the derived spectrum is in good agreement with
measurements of other air Cerenkov telescopes
\citep{1998ApJ...503..744H,2004ApJ...614..897A,1998ApJ...492L..33T,2006A&A...457..899A}.
At energies $<400\,$GeV, below the threshold of previous
measurements by IACTs, we compare our results with those obtained
by CELESTE \citep{2002ApJ...566..343D,2006A&A...459..453S} and
STACEE \citep{2001ApJ...547..949O}, i.e.~measurements performed by
converted solar tower experiments. It should be noted that the
integral flux values of these experiments had to be converted to
differential ones by assuming a shape of the source spectrum,
which causes an additional bias.

\begin{figure}[t]
        \centering
                    \includegraphics*[width=1.08\columnwidth]{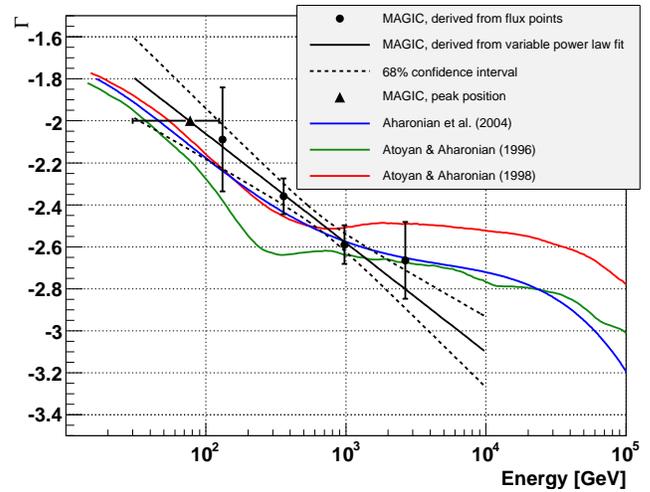}
                        \caption{Measured spectral index derived from differential flux points (\emph{filled circles}) and from the curved power-law
                        fit (\emph{black solid line}; the dashed line gives the $1\sigma$ confidence interval); Predictions
                        by \cite{2004ApJ...614..897A} (\emph{blue line}),
                        \cite{1996MNRAS.278..525A} (\emph{green line}), and
                        \cite{1998nspt.conf..439A} (\emph{red line}) are also shown.
                        }
                        \label{photindex}
  \end{figure}

Above $1\,$TeV up to about 10\,TeV the measured energy spectrum is
well described by a pure  power-law
\citep{1989ApJ...342..379W,2004ApJ...614..897A}. Going to lower
energies, one expects a continuous hardening of the spectrum.
However, this could not be demonstrated by earlier measurements.
The  change of the slope of the spectrum (spectral index)
$\Gamma'$  was tested for various points of the measured spectrum
$\Gamma'$.
\begin{eqnarray}
\Gamma'(E)&=&\frac{d\,\ln(F)}{d\,\ln(E)}\approx\frac{\Delta\,\ln(F)}{\Delta\,\ln(E)}\approx\frac{\ln F_i-\ln F_j}{\ln E_i-\ln E_j}\\
E&=&\exp\left[0.5(\ln E_i + \ln E_j)\right],
\end{eqnarray}
where $F_{i,j}$ is the differential flux measured at $E_{i,j}$.
The four derived spectral indices at $\sim150\,$GeV,
$\sim300\,$GeV, $\sim1\,$TeV, and $\sim2.5\,$TeV shown in Figure
\ref{photindex} indicate a clear softening of the spectrum with
increasing energy. The spectral index $\Gamma'$ was also derived
from the aforementioned results of the curved power-law fit,
\begin{equation}\label{gamma_fit}
\Gamma'=a+2\,b\,\log_{10}\left(E/300\,\mathrm{GeV}\right)
\end{equation}
and is shown by the solid black line, and the
$\pm1\sigma$-confidence band is shown by the dashed black line. A
systematic uncertainty on the slope can cause an additional
vertical shift of the measurement by $\pm0.2$. Within
uncertainties, the measured spectral index varies in good
agreement with predictions by \cite{2004ApJ...614..897A} (\emph{blue line}), who, in addition to the IC-scattering on synchrotron
photons, included several other soft photon fields such as
millimeter-photons, CMB, and far-IR photons from dust and stars.

The predicted GeV $\gamma$-ray emission has a peak in the
SED-representation (see Figure \ref{Crab_sed_time}). If one
assumes that the energy spectrum around the peak can be described
with a curved power-law, the position of the peak can be
determined from the measurement of the spectral index obtained
from the result of the curved power-law fit. A necessary condition
for the peak in the SED is that the spectral index $\Gamma'$ is
$-2$. With this condition the peak is determined at
$77\pm47_\mathrm{stat}{+107\atop -46}_\mathrm{syst}$\,GeV
(Figure \ref{photindex}, \emph{triangle}).

\begin{table}[b]

\def\arraystretch{1.5}
    \centering
        \caption{\label{diffpoints}Mean Energy and Differential Flux of the Spectral Points Shown in Figure \ref{crab:diffspectr}}
       \begin{tabular}{cc}\hline\hline
        \rule[0mm]{0mm}{ 3mm}{Energy}{Mean Energy} & {Differential Flux}\\

         \rule[-2mm]{0mm}{ 0mm}(GeV)   & (cm$^{-2}$s$^{-1}$TeV$^{-1}$) \\
        \tableline
          77........................................ & $(1.14\pm0.27_\mathrm{stat}\pm0.34_\mathrm{syst})10^{-8}$\\
          127...................................... & $(3.65\pm0.38_\mathrm{stat}\pm0.55_\mathrm{syst})10^{-9}$\\
            210...................................... & $(1.41\pm0.09_\mathrm{stat}\pm0.28_\mathrm{syst})10^{-9}$\\
        346...................................... & $(4.37\pm0.23_\mathrm{stat}\pm0.87_\mathrm{syst})10^{-10}$\\
        570...................................... & $(1.32\pm0.07_\mathrm{stat}\pm0.20_\mathrm{syst})10^{-10}$\\
        940...................................... & $(3.55\pm0.23_\mathrm{stat}\pm0.18_\mathrm{syst})10^{-11}$\\
    1550.................................... & $(9.88\pm0.74_\mathrm{stat}\pm0.49_\mathrm{syst})10^{-12}$\\
    2554.................................... & $(2.69\pm0.29_\mathrm{stat}\pm0.27_\mathrm{syst})10^{-12}$\\
    4212.................................... & $(6.80\pm1.10_\mathrm{stat}\pm1.00_\mathrm{syst})10^{-13}$\\
    6943.................................... & $(1.15\pm0.53_\mathrm{stat}\pm0.12_\mathrm{syst})10^{-13}$\\\tableline
       \end{tabular}
       \tablecomments{The systematic errors are derived from different applied cuts and unfolding procedures.}
    \end{table}

\subsection{VHE $\gamma$-Ray Light Curve of the Nebula Emission}

In the VHE $\gamma$-ray astronomy community it is assumed that the
Crab Nebula is a constant and stable $\gamma$-ray source and can
therefore be used as a standard candle. However, with more
sensitive measurements it is necessary to check the stability of
the $\gamma$-ray source. Below we present a time-resolved
measurement of the VHE $\gamma$-ray flux, i.e.~the light curve for
the Crab Nebula. Depending on the source strength and the analysis
threshold, the time intervals can be as short as a few minutes.

\begin{figure*}[htb]
        \centering
         \includegraphics*[bb = 25 33 541 747, angle=0,
                    width=0.78\textwidth]{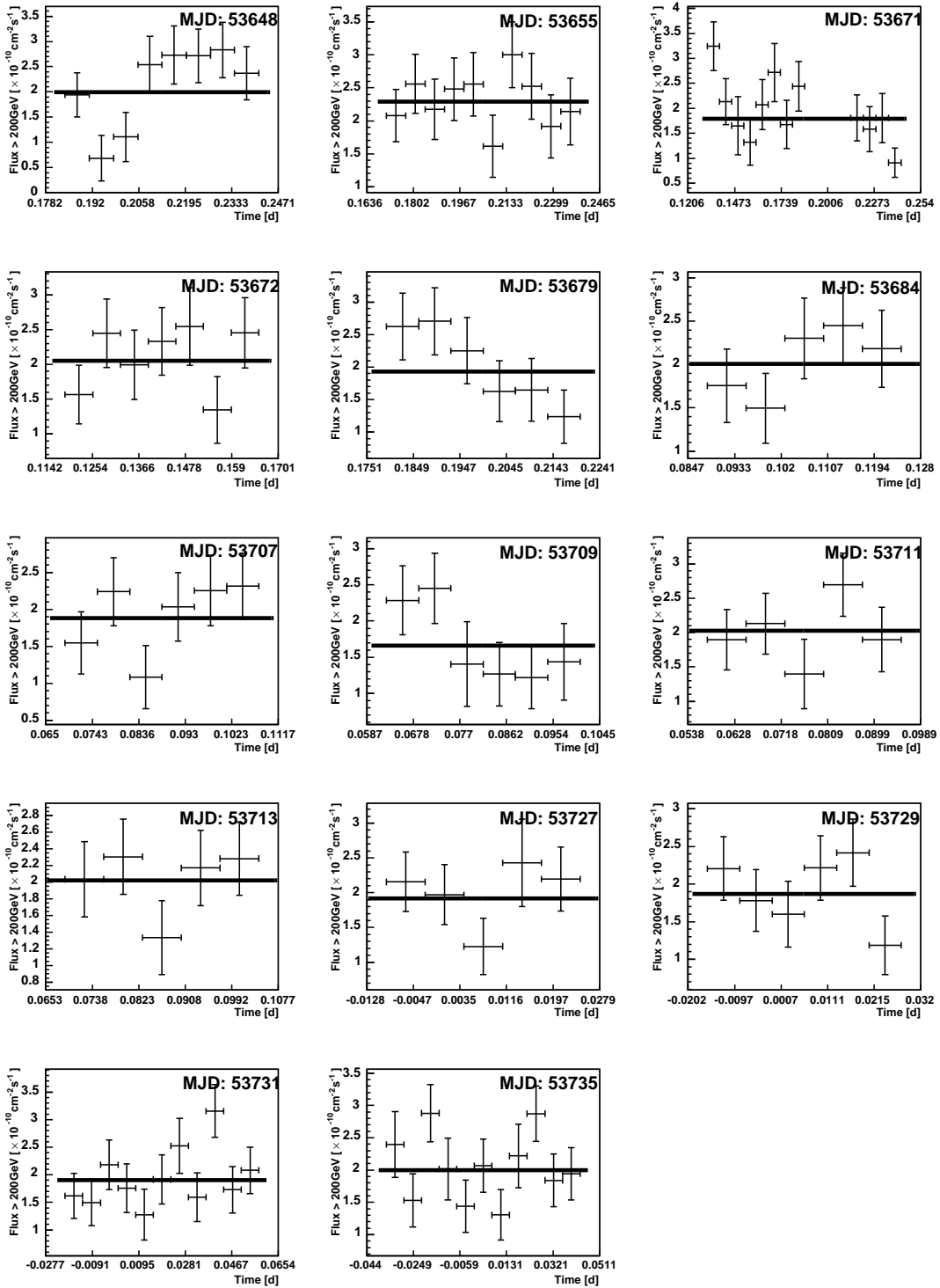}
                        \caption{ Light curves of the integral Flux above 200 GeV from the Crab Nebula for each night.}
                        \label{crab:lcbyday}
  \end{figure*}

\begin{figure}[t]
        \centering
         \includegraphics*[bb = 53 41 415 701, angle=0,
                    width=\columnwidth]{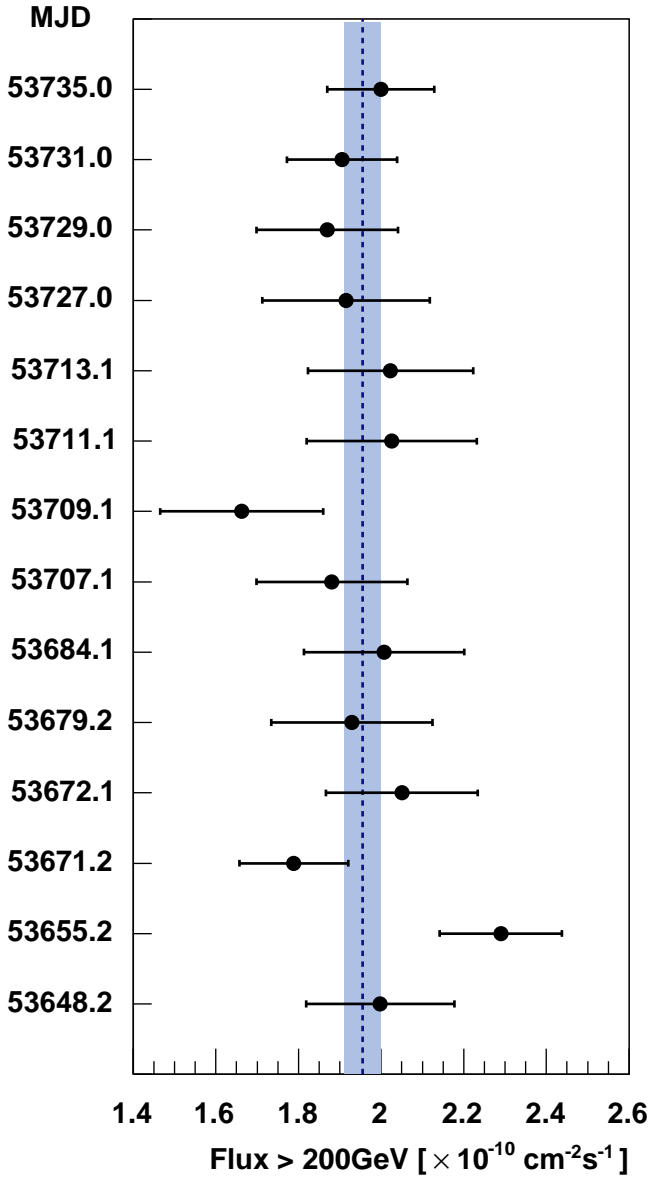}
                        \caption{Nightly average flux from the Crab Nebula above 200GeV of each observed night.
                        The dashed blue line gives the average flux of all
                        nights
                         and the blue shaded region gives the corresponding statistical error.}
                        \label{crab:lcsummary}
  \end{figure}

We calculated light curves in bins of 10 minutes from events with
estimated energies above $200\,$GeV. The light curves of all 14
selected nights are shown in Figure \ref{crab:lcbyday}. Note that
the same loose cuts are used for the $\gamma$/hadron-separation as
for the calculation of the energy spectrum, which reduces the
sensitivity of the measurement. The probability that the light
curve is described by a constant flux level is $>10\%$ in all
nights except the first night, where the probability of the fit is
$0.8\%$. The average statistical uncertainty of each flux
measurement is $\sim20\%$. Figure \ref{crab:lcsummary} shows the
average flux of each night. The dashed line in the figure denotes
the average flux from all nights, and the shaded region shows the
statistical error in the flux. The average integral mean flux
$F_{>200\,\mathrm{GeV}}$ is
\begin{equation}
F_{>200\,\mathrm{GeV}}=(1.96\pm0.05_{\mathrm{stat}})\times
10^{-10}\,\mathrm{cm}^{-2}\,\mbox{s}^{-1}\quad.
\end{equation}
There is a probability of 67\% that the measured daily flux values
are compatible with a constant flux. We can, therefore, conclude
that the reconstructed flux of the Crab Nebula, within statistical
uncertainties, was constant over the entire observation period.

\subsection{Morphology of the $\gamma$-Ray emitting Region}

The morphology of the $\gamma$-ray emission was studied by
generating sky-maps in three uncorrelated bins of SIZE.
The reconstruction of the origin of a $\gamma$-ray event with a
single telescope is possible with the so-called
DISP-method
\citep{1994APh.....2..137F,2001APh....15....1L}. For the studies
presented here we used the following parameterization for
DISP:
\begin{equation}
\mbox{DISP} =
a\left(\mbox{SIZE}\right)+b\left(\mbox{SIZE}\right)\cdot\frac{\mbox{WIDTH}}
    {\mbox{LENGTH}}\quad,
\end{equation}
where $a$ and $b$ are second-order polynomials found by fitting
MC simulated $\gamma$-ray showers
\citep{2005ICRC....5..363D}. Strong tail cuts of 10 and 8
phe were used in the image cleaning for core- and
boundary-pixels, respectively, and a tight HADRONNESS cut
$<0.1$ was applied, resulting in improved angular resolution.

The reconstructed event origins were corrected for possible
mispointing by using the information from the starguider camera.
Two-dimensional (2D) histograms with bin sizes of $(0.057^\circ\times0.057^\circ)$
were filled with the corrected event origins (events with energies
$<500\,$GeV). A 4 times finer binning was chosen for the
sky-map filled by events with energies above $500\,$GeV.  Figure
\ref{disp_skymaps} shows the background-subtracted sky-maps of
excess events from the Crab Nebula for $\gamma$-ray energies
$\sim160$, $\sim250$, and $>500\,$GeV.

\subsubsection{Center of Gravity of the $\gamma$-Ray Emission}

\begin{figure*}[t]
        \centering
        \mbox{}
\includegraphics*[width=1.\columnwidth]{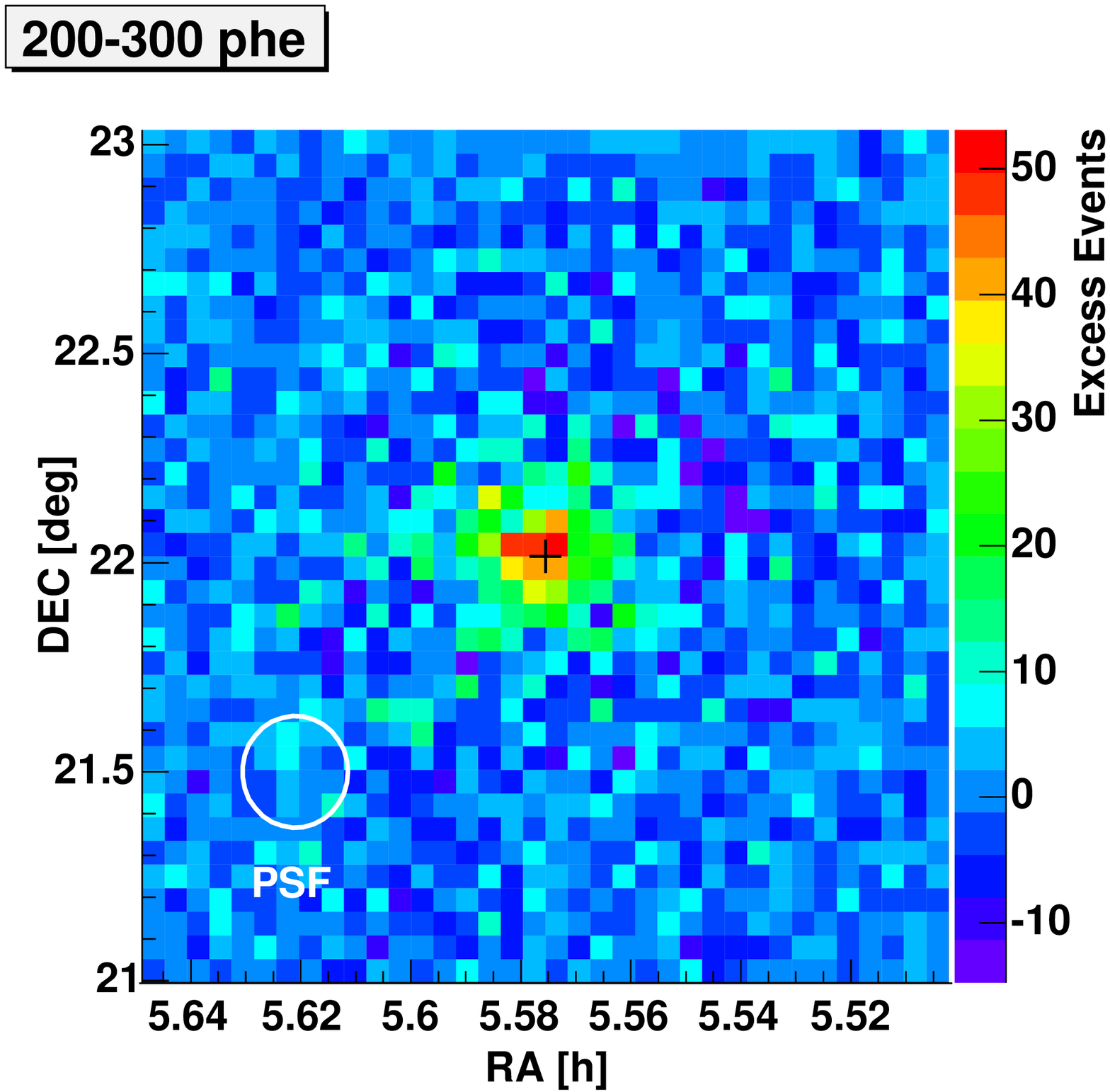}
\includegraphics*[width=1.\columnwidth]{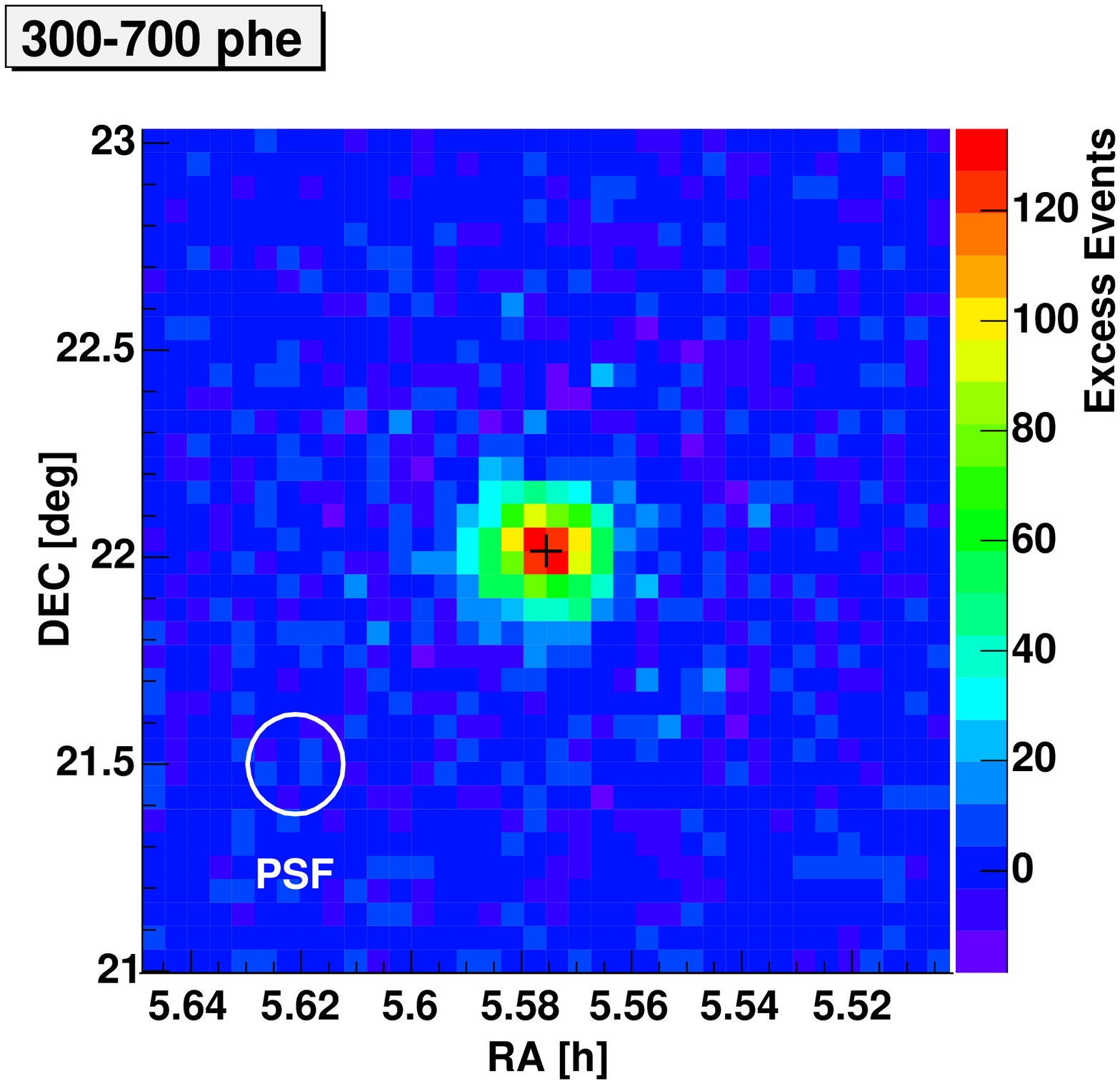}
\includegraphics*[width=1.\columnwidth]{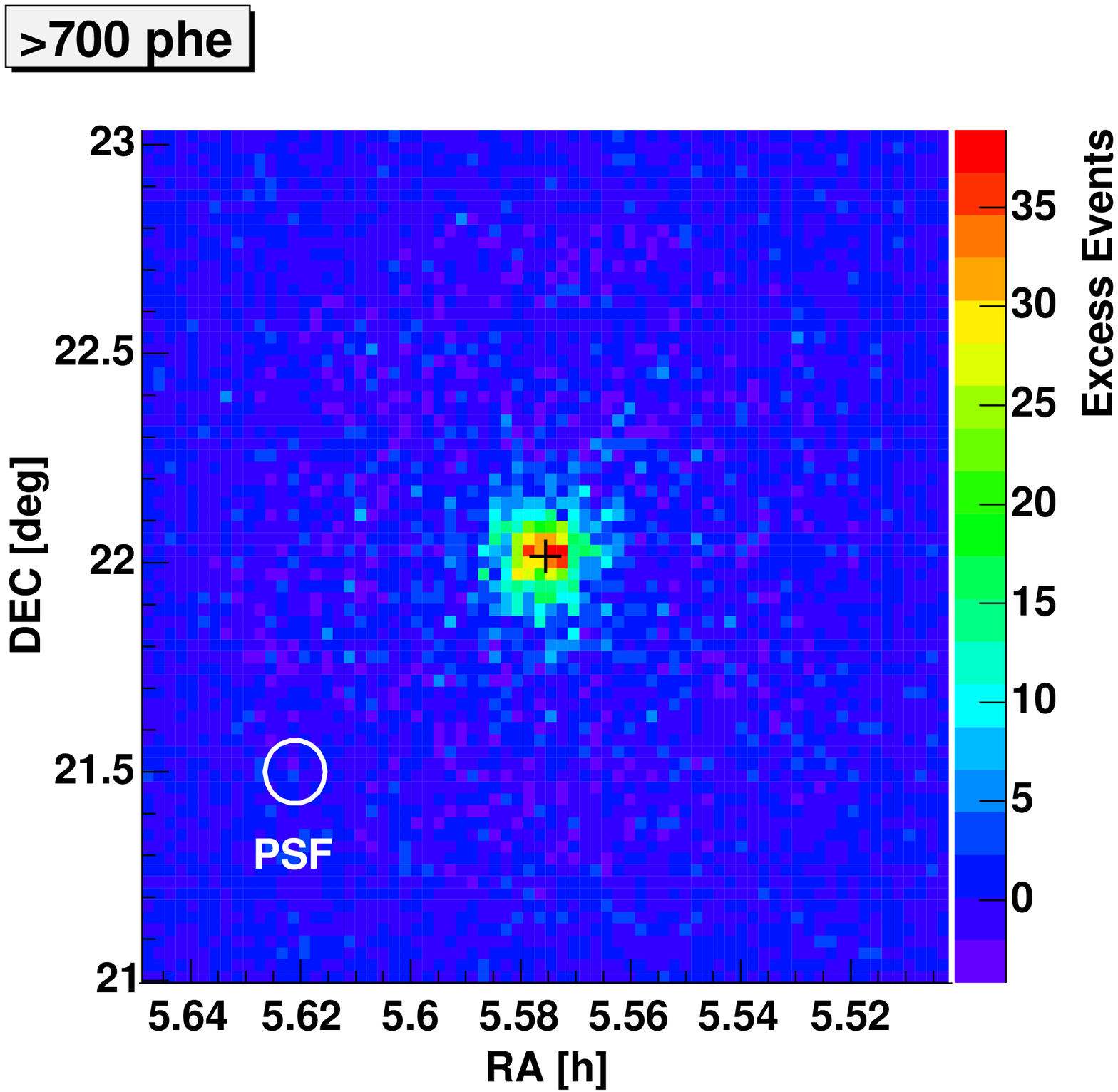}
\caption{Sky-maps of excess events from the Crab Nebula for
different $\gamma$-ray energies ($\sim160$, $\sim250$,
and $>500\,$GeV). The position of the pulsar is marked by the
black plus sign and the angular resolution is indicated by the
circle.}
\label{disp_skymaps}
  \end{figure*}

The CoG of the $\gamma$-ray emission was
derived from the sky-maps of the excess events shown in Figure
\ref{disp_skymaps} by fitting them with a 2D-Gaussian of the form
\begin{equation}\label{2dgaus}
    F(x,y) =
    F_{\mathrm{res}}+a\cdot\exp\left[-\frac{(x-\bar{x})^2+(y-\bar{y})^2}{2\sigma^2}\right]\quad,
\end{equation}
where $F_{\mathrm{res}}$ is introduced to account for a possible
constant offset of the background-subtracted sky-map. In this
representation $\sigma$ defines the 39\% containment radius of the
observed $\gamma$-ray emission. Here we assume that the
distribution of excess events is rotationally symmetric,
i.e.~$\sigma_x=\sigma_y=\sigma$.  It is further assumed that
$\sigma$ is the convolution of the response of the detector
$\sigma_{\mathrm{psf}}$ (point-spread function) and the apparent
size of the $\gamma$-ray emission region $\sigma_{\mathrm{src}}$,
i.e.~$\sigma^2=\sigma_{\mathrm{psf}}^2+\sigma_{\mathrm{src}}^2$.
Note that here it is assumed that the $\gamma$-ray emission region
follows the shape described by Equation \ref{2dgaus}, which in
reality is not necessarily the case.

The CoGs obtained from the fitted $\bar{x}$ and $\bar{y}$ are
listed in Table \ref{crabpositions} and shown in Figure
\ref{crabcomposite} superimposed on the composite image of
optical, IR, and X-ray observations of the Crab Nebula. The three
measured CoGs are compatible among each other and coincide with
the position of the pulsar. Note that the systematic uncertainty
of the position is $\sim1\arcmin$.

\subsubsection{Extension of the $\gamma$-Ray Emission Region}

The extension of the $\gamma$-ray emission region was studied by
comparing the width of the excess event distribution with that
obtained for a simulated $\gamma$-ray point-source. The simulated
distributions were verified by comparing them to the distributions
extracted from an observation of Mrk 421, an extragalactic
$\gamma$-ray source that can be considered a point source for our
purpose. The Mrk 421 data set is the same as in
\cite{albert:2007:magic:mkn421}. The width of the $\gamma$-ray
excess extracted from Mrk 421 and the simulated width for a point
source agree within statistical uncertainties.

\begin{figure*}[t]
        \centering
         \includegraphics*[width=1.03\columnwidth]{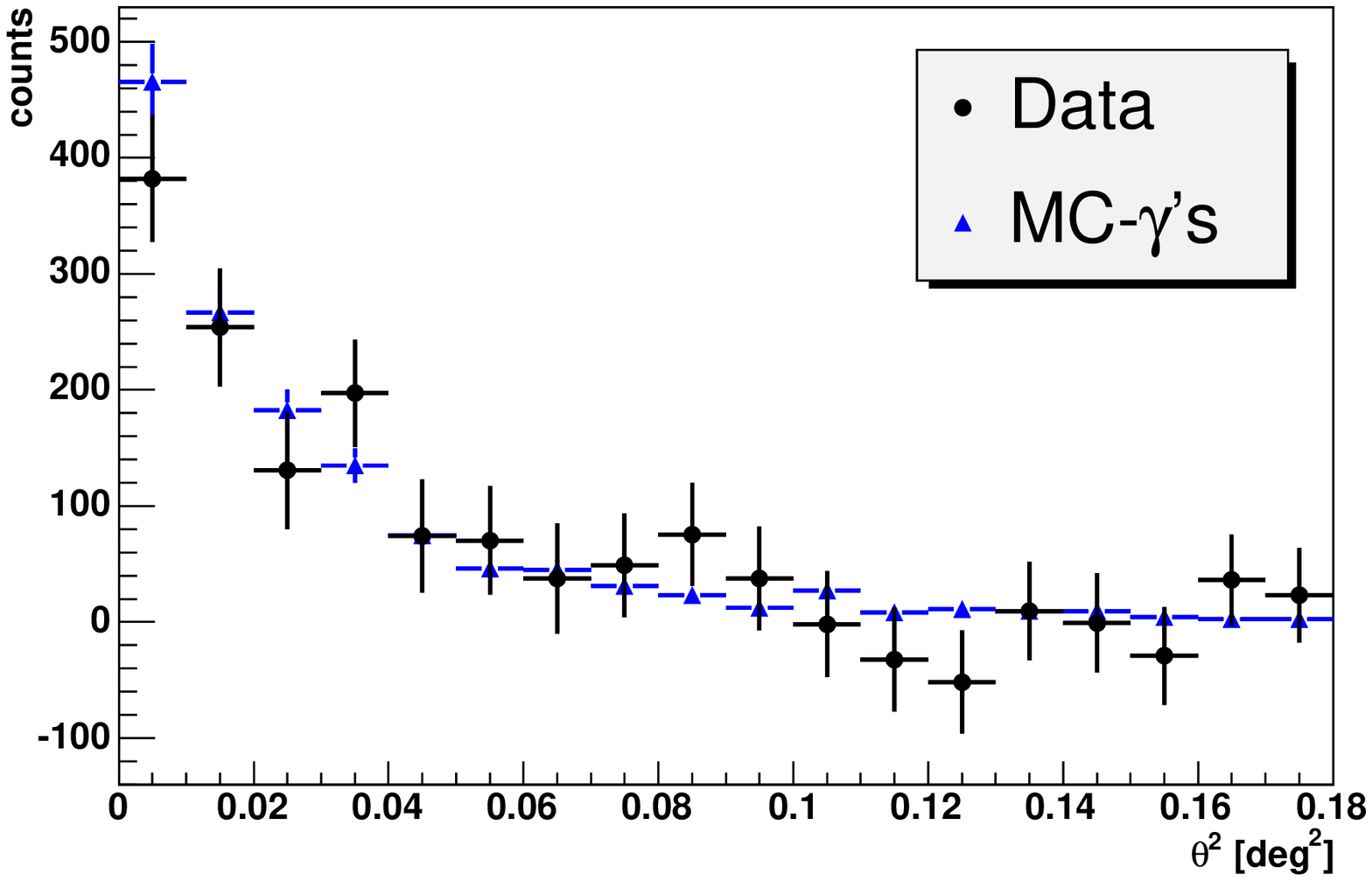}
         \includegraphics*[width=1.03\columnwidth]{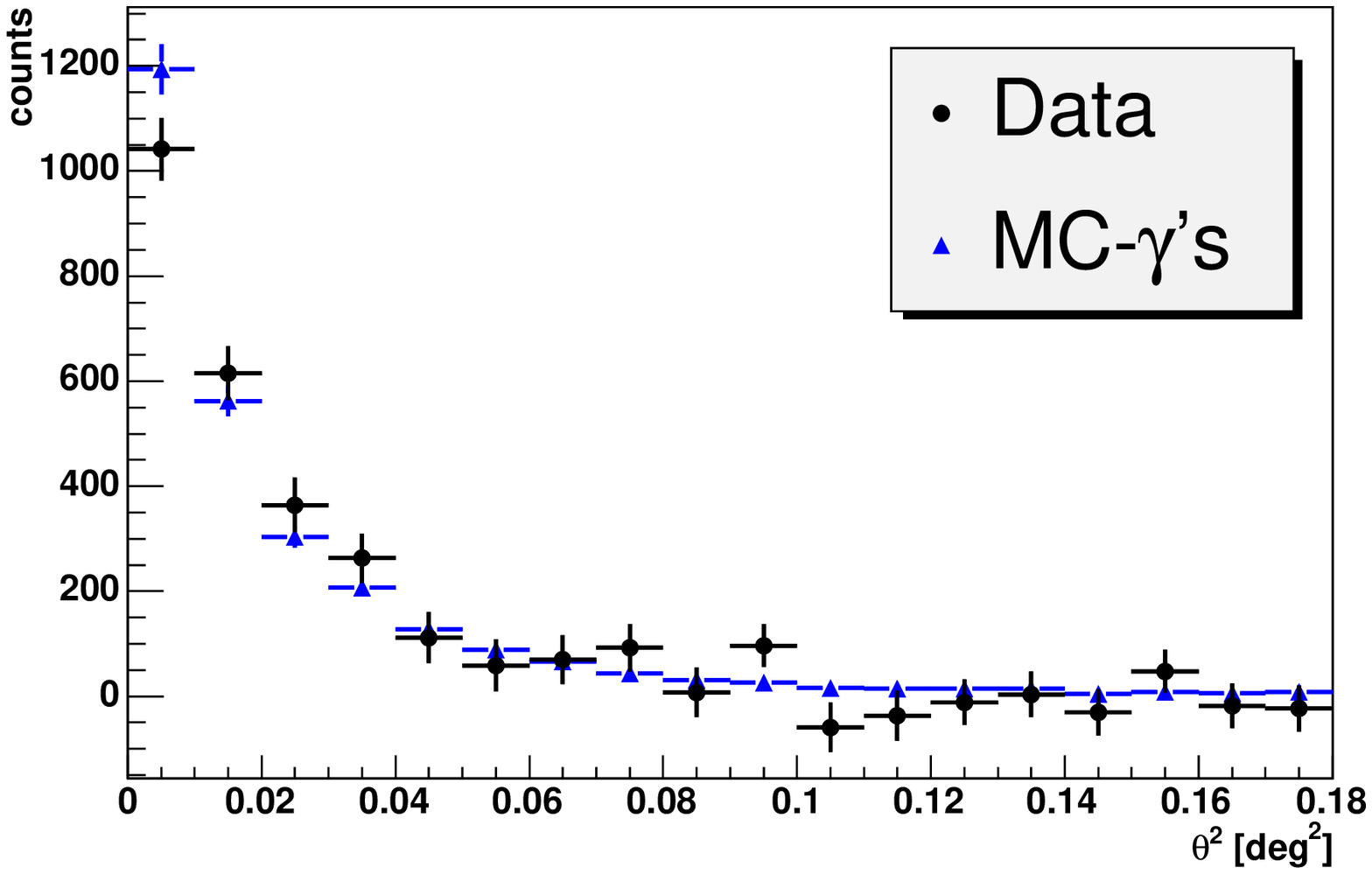}\\
        \includegraphics*[width=1.03\columnwidth]{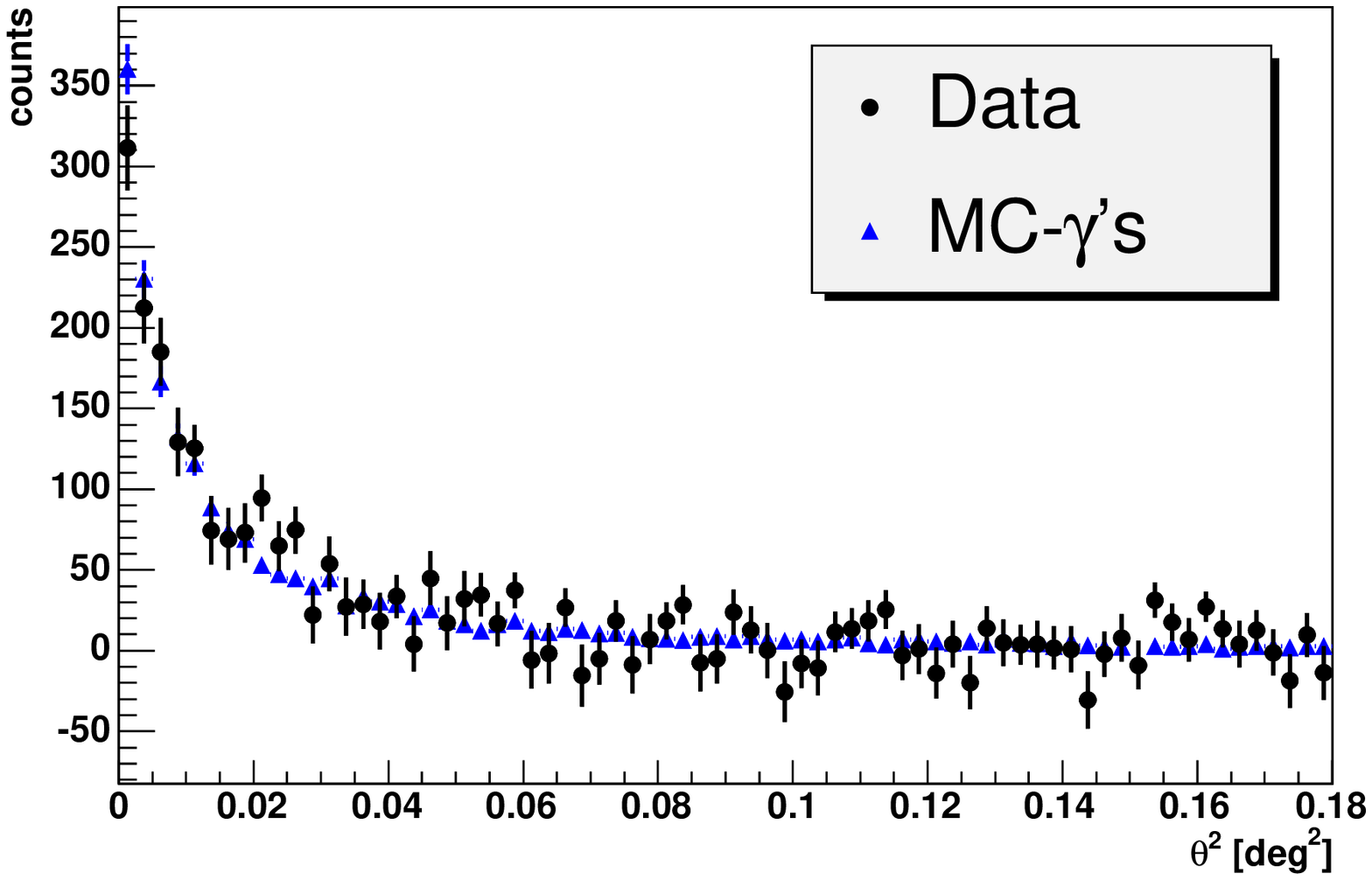}
                        \caption{Background subtracted $\theta^2$-distributions for different
                        energies. $\sim160\,$GeV \emph{(top left)}, $\sim250\,$GeV \emph{(top right)}, $>500\,$GeV \emph{(bottom)}
                        }
                        \label{disp_theta2maps}
  \end{figure*}

In the following, the average position of the CoGs obtained from
the three sky-maps is assumed as the $\gamma$-ray source position.
The angular distance squared ($\theta^2$) between the
reconstructed origin and the assumed source position is calculated
for every event. The background subtracted
$\theta^2$-distributions obtained for the three energy ranges are
shown in Figure \ref{disp_theta2maps}. Data (\emph{black}) and MC (\emph{blue})
are compatible within statistical uncertainties in all three
$\theta^2$-distributions.

An exponential function of the form
\begin{equation}\label{theta2}
F\left(\theta^2\right) = a \cdot
    \exp\left(-\frac{\theta^2}{2\sigma^2}\right)
\end{equation}
describes the expected $\theta^2$-distribution, where $a$ is a
normalization and $\sigma$ is the same as in equation
\ref{2dgaus}. Values for $\sigma^2$ and $\sigma^2_\mathrm{psf}$
found by fitting the corresponding $\theta^2$-distributions with
equation \ref{theta2} are shown in Table
\ref{crabextensionexcess}. Upper limits on $\sigma_{\mathrm{src}}$
were calculated with a confidence level of 95\% following the
procedure outlined in \cite{PDBook} for one-sided confidence
intervals and Gaussian errors. The results are presented in Table
\ref{crabextensionexcess}. For energies above $500\,$GeV the limit
is shown  in Figure \ref{crabcomposite}. The limits obtained for
$\gamma$-ray energies above $500\,$GeV and about $250\,$GeV
constrain the $\gamma$-ray emission to a region within the optical
synchrotron nebula.

  \subsection{Search for pulsed $\gamma$-Ray Emission}\label{ac}

Among the most challenging tasks of ground-based $\gamma$-ray
experiments is the detection of a pulsar. Several experiments have
tried but failed. Currently MAGIC is the only ground-based
detector with threshold settings below 100\,GeV that is
appropriate for a search of pulsed $\gamma$-ray emission from the
Crab pulsar. For the data a periodicity analysis was performed
after $\gamma$/hadron-separation and selection of events with
small $|\mbox{ALPHA}|$-value. The cuts were chosen by MC simulations to optimize the sensitivity of the analysis.
After event selection, the event times\footnote{The time of each
event was derived from the time signal of a GPS controlled
rubidium clock with a precision of $\sim200$ ns.} were
transformed to the barycenter of the solar system with the TEMPO
timing package \citep[][\footnote{Available at http://www.atnf.csiro.au/research/pulsar/tempo/.}]{tempo}. Then, the corresponding phase
$\phi_j$ of the Crab pulsar was calculated for each transformed
arrival time $t_j$:
\begin{equation}
\phi_j=\nu(t_j-t_0)+\frac 1 2 \dot{\nu}(t_j-t_0)^2\quad,
\end{equation}
where $\nu$, $\dot{\nu}$, and $t_0$ are values of contemporary
ephemerides of the Crab pulsar provided by the Jodrell Bank Radio Telescope\footnote{See http://www.jb.mac.ac.uk/pulsar/crab.html.}
(see Table \ref{ephemeris}).  We tested for periodicity  with the
\emph{H}-test \citep{1989A&A...221..180D}, the Pearson's $\chi^2$-test,
and a test from \cite{1992ApJ...398..146G} that is based on
Bayesian statistics.

\begin{table}[b]
    \centering
    \def\arraystretch{1.5}
        \caption{\label{crabpositions}Center of Gravity
        of the $\gamma$-Ray Emission of the Crab
        Nebula Obtained for Different Energies}

       \begin{tabular}{cccc}\hline\hline
         \rule[0mm]{0mm}{ 3mm}SIZE & Energy  & Right
        Ascension & Declination\\
        \rule[-2mm]{0mm}{ 0mm}\mbox{(phe)} & (GeV) & (hr)     & (deg) \\\hline
 $200-300$&$160^{+80}_{-50}$  & $5.5766\pm0.0009$ & $22.019\pm0.011$ \\
 $300-700$&$250^{+130}_{-80}$  & $5.5758\pm0.0003$ & $22.019\pm0.004$ \\
 $>700$& $>500$ &$5.5759\pm0.0003$ & $22.022\pm0.003$\\
 Position of the pulsar: & \ldots & 5.5755 &  22.015   \\\hline
       \end{tabular}
       \tablecomments{In the first column the applied SIZE cut is stated.
         The second column
        shows the corresponding range of $\gamma$-ray energies covered (peak value and full width at half maximum of the
        distribution of MC $\gamma$-ray events). The last two columns give the fitted position of the
        CoG and the statistical uncertainty.}
    \end{table}

\begin{figure}[b]
        \centering
                    \includegraphics*[bb = 200 52 560 720, angle=-90,width=1.\columnwidth]{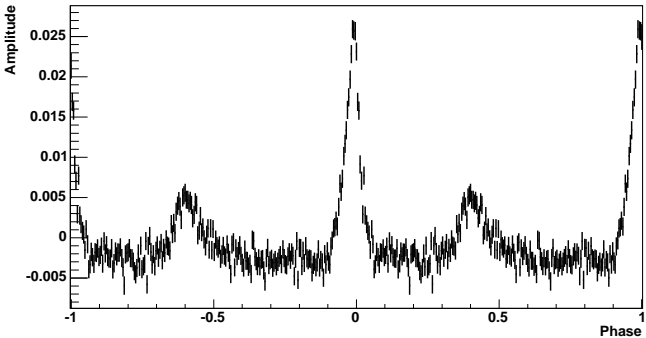}
                        \caption{Optical light curve of the Crab pulsar
                        measured with MAGIC. The figure includes data from
                        seven different observations between 2005 December and 2006 February.
                        The total observation time was 12.5 hr.}
                        \label{Crab_optical_light_curve}
   \end{figure}

The analysis chain was tested by optical observations of the Crab
pulsar with the MAGIC telescope. Within this $12.5\,$hr
observation, every time the readout of MAGIC was triggered by a
cosmic-ray shower, the signal of the pixel in the center of the
MAGIC camera was recorded by the MAGIC DAQ for 100\,ns. Along with
an average trigger rate of 200\,Hz, the effective observation time
was only about 1 s. Figure \ref{Crab_optical_light_curve}
shows the reconstructed optical light curve of the Crab pulsar
with the familiar main pulse and interpulse. For better readability
the light curve is shown twice. The position of the main-pulse is
shifted with respect to the position of the main-pulse in radio by
$-252\pm64\mu$s, which is in agreement with the contemporary
measurement of \cite{2006astro.ph..6146O}.

\subsubsection{Search for pulsed Emission in Differential Bins of Energy}

We searched for pulsed $\gamma$-ray emission in five bins of
reconstructed energy between $60\,$GeV and $9\,$TeV. This search
was motivated by a possible pulsed $\gamma$-ray component at TeV
energies \citep{2001ApJ...549..495H,2007astro.ph..1676H}. However,
no signature of periodicity was found in any of the tested energy
intervals.

\begin{figure}[t]
        \centering
\includegraphics*[bb= 120 28 723 584, width=0.9\columnwidth]{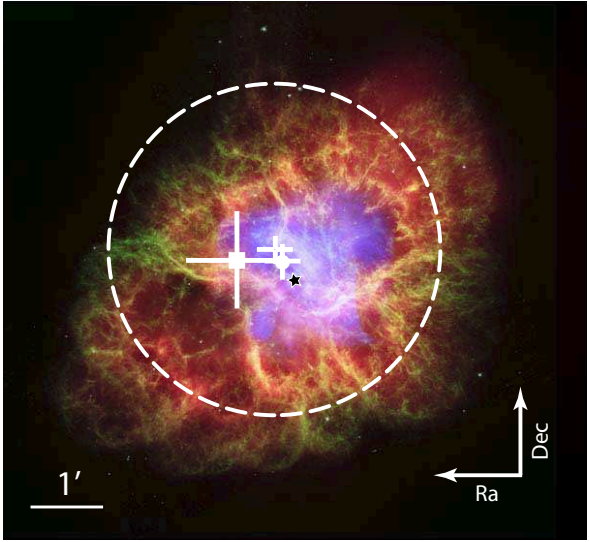}
                        \caption{Emission of the Crab Nebula in different
                          energy bands. The position of the Crab pulsar is marked with a black star.
                          The \emph{Chandra} X-ray image is shown in light blue,
                           the \emph{Hubble Space Telescope} optical images are in green and
                            dark blue, and the S\emph{pitzer Space Telescope}'s infrared image
                             is in red \citep[picture from][]{chandra_crabcomp}.
                             Overlaid are the CoG of the $\gamma$-ray emission at different energies
                             (\emph{plus sign:} $>500\,$GeV; \emph{filled circle:} $\sim250\,$GeV;
                             \emph{filled square:} $\sim160\,$GeV). The error bars indicate the statistical uncertainty
                             in the position of the CoG.
                             Indicated by the dashed circle is the upper limit (95\% confidence level) on the 39\% containment radius of the $\gamma$-ray emission region that was derived from the
                             $\theta^2$-distribution for $\gamma$-ray energies above
                             500\,GeV.}
                        \label{crabcomposite}
  \end{figure}

\begin{table}[b]
    \centering
        \caption{\label{crabextensionexcess}Results of the fit of the
        $\theta^2$-distributions with an exponential ansatz and thereof derived
        upper limits on the extension of the emission region (39\% containment radius). }
     \begin{tabular}{cccc}\hline\hline

        \rule[0mm]{0mm}{ 3mm}Energy&Data $\sigma^2$&MC $\sigma_\mathrm{psf}^2$&95\% U.L. on Extension\\
         \rule[-2mm]{0mm}{ 0mm}(GeV) & (deg$^2$)&
                      (deg$^2$) &(arcmin)\\\hline

                      $160^{+80}_{-50}$&$0.0148\pm0.0035$&$0.0113\pm0.0007$  &5.9\\
  $250^{+130}_{-80}$&$0.0100\pm0.0008$ &$0.0100\pm0.0003$& 2.4\\
 $>500$ &$0.0054\pm0.0006$ &$0.0051\pm0.0002$& 2.2    \\\tableline
       \end{tabular}

    \end{table}
For each energy bin an upper limit on the number of excess events
was calculated with a confidence level of 95\% in two different
ways, first from the result of the \emph{H}-test as described by
\cite{1994ApJ...436..239D}, and second from the pulse phase
profile. In the calculation of the limit from the result of the
\emph{H}-test, it is assumed that the duty cycle (FWHM) of the pulsed $\gamma$-emission is 20\%, similar to the
duty cycle of the light curve measured  by the EGRET detector
above 100\,MeV \citep{1998ApJ...494..734F}. No assumption about
the position of the emission in the pulse phase profile enters
into the calculation.

This additional constraint is applied,
however, when the upper limit is directly derived from the pulse
phase profile. As signal regions we chose the phase intervals
where EGRET had observed pulsed emission above 100\,MeV,
i.e.~$-0.06--0.04$ and $0.32--0.43$ (shaded region in
Figure \ref{lcoptcuts} below). The background was estimated from the
remaining phase intervals. Having defined the signal and
background regions in this way, the upper limit on the number of
excess events was obtained by the method of
\cite{2005NIMPA.551..493R}. Because of the additional constraint
made about the position of the expected pulsed emission, the
limits obtained from the pulse phase profile are on average about
a factor of 2 better than the limits obtained from the result of
the \emph{H}-test.

\begin{figure}[b]
        \centering
\includegraphics*[width=\columnwidth]{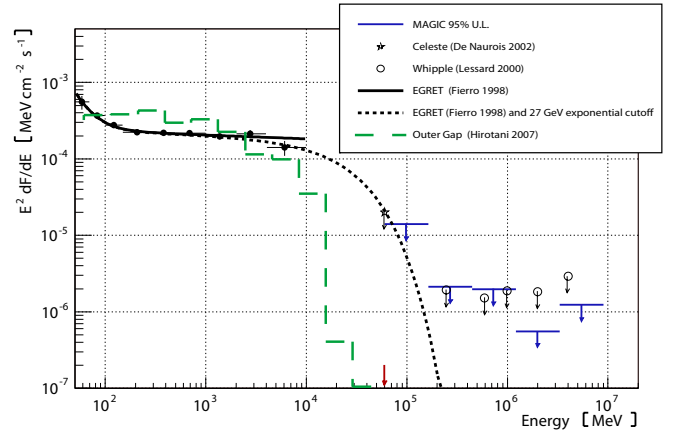}
                        \caption{Upper limits on the pulsed $\gamma$-ray flux from the Crab
                        pulsar; upper limits in differential bins of energy are given
                        by the blue points. The upper limit on the
                        cutoff energy of the pulsed emission is
                        indicated by the dashed line. The
                        analysis threshold to derive the upper
                        limit on the cutoff energy is indicated by the red arrow.
                        }
                        \label{Crab_pulsed_differential}
    \end{figure}

The upper limits derived from the pulse phase profiles were
converted into flux limits. The collection area was calculated
assuming a photon index of -2.6 for the $\gamma$-ray spectrum. The
flux limits are shown in Figure \ref{Crab_pulsed_differential}
together with the upper limit on the cutoff energy, which is
derived in the following section.

\begin{table}[t]
    \centering
        \caption{\label{ephemeris}Ephemerides of the Crab Pulsar from the Jodrell Bank Telescope
        Covering the Same Period of Time as the Analyzed Data}
       \begin{tabular}{cccc}\hline\hline
         &  {${t}$} & {Frequency ($ \nu$)} & {Derivative ($\dot{\nu}$)}\\
 JD& (s)& (Hz) & ($10^{-15}$ Hz s$^{-1}$)\\\tableline

2,453,597.5............... &  0.029626  &   29.7798524524  & -372992.36 \\
2,453,628.5............... & 0.031767 &     29.7788534525  & -372972.07 \\
2,453,658.5............... & 0.022656  &    29.7778867428  & -372950.45 \\
2,453,689.5............... & 0.016803  &    29.7768878849  &-372924.54  \\
2,453,719.5............... & 0.026788   &   29.7759213143  &-372886.52 \\
2,453,750.5............... & 0.020341  & 29.7749226318    & -372854.62 \\
2,453,781.5............... & 0.006520  &29.7739240139     &-372823.70
\\\tableline

       \end{tabular}
       \tablecomments{Given in JD is the reference day of the ephemeris, and $t$ is the time of appearance
        of the first main pulse on the reference day after midnight.}
    \end{table}

    \subsubsection{Upper Limit on the Cutoff Energy of the pulsed Emission}

Apart from the search for pulsed emission in bins of reconstructed
energy, we performed a periodicity analysis, this time selecting
events with SIZE$<300\,$phe ($\gamma$-ray energies
$\lesssim180\,$GeV) and applying the same optimized SIZE-dependent HADRONNESS cuts and ALPHA cuts as
above. Compared to the previously described analysis, this one is
optimized for a search of pulsed emission close to the threshold
of the experiment. The analysis threshold, defined as the peak of
the energy-distribution of simulated $\gamma$-ray showers, is
$60\,$GeV.

\begin{figure}[b]
   \centering
\includegraphics*[width=\columnwidth]{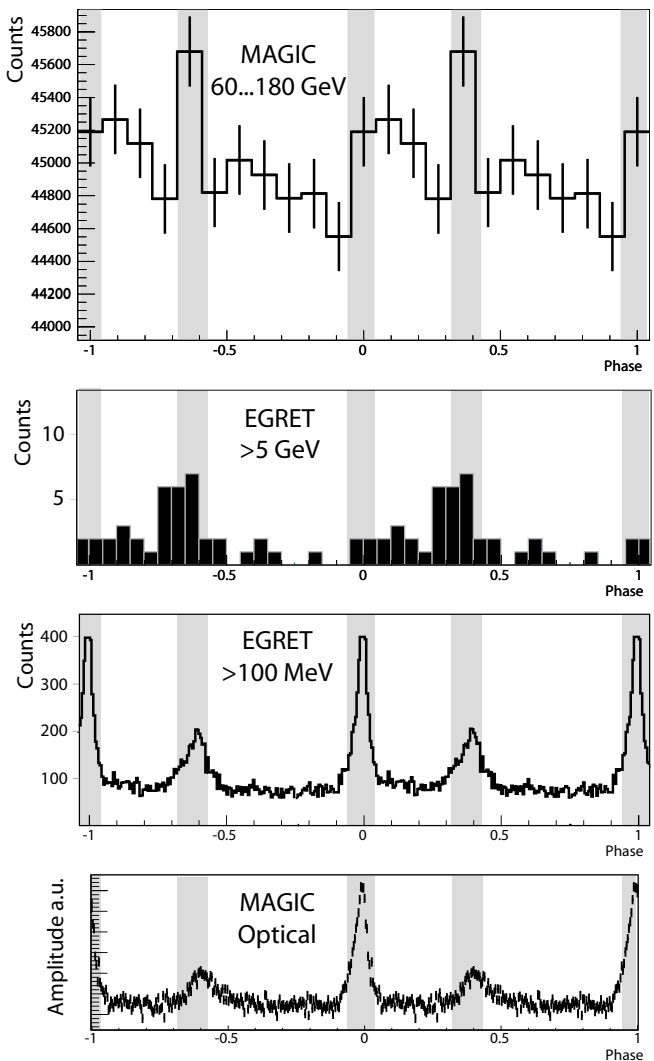}
\caption{\label{lcoptcuts}Pulse phase profiles of the Crab pulsar.
\emph{Bottom panel}: optical observations by MAGIC ($\sim1\,$s effective
observation time); \emph{middle panels}: observations by EGRET from
\cite{2004ASSL..304..149T}; \emph{top panel}: pulse phase profile
obtained by MAGIC, analysis threshold 60\,GeV. The shaded regions
indicate the EGRET measured positions of the peaks for
$\gamma$-ray energies above 100\,MeV. Note that each pulse phase
profile is shown twice for better visibility.}
   \end{figure}

Figure \ref{lcoptcuts} shows the pulse phase profile obtained for
the selected events. For comparison the pulse phase profiles from
EGRET observations above 100\,MeV and 5\,GeV
\citep{2004ASSL..304..149T} are also shown. The EGRET pulse phase
profile above 10\,GeV \citep{2005ApJS..157..324T} is not shown
because it suffers from too low statistics. Shaded in the pulse
phase profiles are the regions of the main and interpulse
defined from EGRET observations above 100 MeV
\citep{1993ApJ...409..697N}.

The result of a Pearson's $\chi^2$-test is $13.1$ with 10 degrees
of freedom, corresponding to a significance of $1.2\,\sigma$ for
periodic emission. The result of the \emph{H}-test is 3.9, which is
equivalent to a significance of $1.3\,\sigma$. The test by
\cite{1992ApJ...398..146G} results in a probability of
$4.1\cdot10^{-4}$ that pulsed emission is present in the data.
These tests do not make an assumption about the position of the
pulsed emission in the pulse phase profile. However, some evidence
of an excess is visible at the position of the inter-pulse in the
same phase range where EGRET detected pulsed emission above
100\,MeV.  If the two phase regions defined by EGRET are used as
the signal region and the remaining phase intervals as background
region, the significance of the observed excess is $2.9\,\sigma$.
Note that in this case the significance was not calculated from
the binned pulse phase profile shown in Figure \ref{lcoptcuts}.

The significance of the observed excess is not sufficient to claim
the detection of a pulsed signal; therefore, upper limits on the
number of excess events were calculated with a confidence level of
95\% (see Table \ref{crab300phe}). Note that because of the
observed excess, the upper limit from the pulse phase profile is
larger than the limit obtained from the \emph{H}-test. Using the
different limits on the number of pulsed excess events, we
constrain, in the following, the cutoff energy of the pulsar
spectrum under the assumption that the break in the energy
spectrum can be described with an exponential cutoff. In the
procedure we use the parametrization of the measured pulsar
spectrum below $10\,$GeV \citep{1998ApJ...494..734F}, extended
with an exponential cutoff:
\begin{eqnarray}\label{crabegret}
    F(E,E_\mathrm{
    Cutoff})&=&\left[7.0\cdot10^{-6}\left(\frac{E}{0.1\,\mathrm{GeV}}\right)^{-4.89}\right. \nonumber\\
    &&\left. +2.3\cdot10^{-5}\left(\frac{E}{0.1\,\mathrm{GeV}}\right)^{-2.05}
    \right]\nonumber\\
    &&\times\exp{\left(-\frac{E}{E_{\mbox{\scriptsize
    Cutoff}}}\right)}\,\mathrm{photons}\left(\mathrm{cm}^{2}\,\mbox{s}\,\mbox{GeV}\right)^{-1}.
\end{eqnarray}

\begin{table*}[t]
    \centering
        \caption{\label{crab300phe}Analysis Results for a Cut
   selecting Events with  SIZE
   $<300\,\mathrm{phe}$}
       \begin{tabular}{ccc}\hline\hline
        \rule[-2mm]{0mm}{ 6mm}{Method} & \emph{H}-Test & Rolke\\\hline
        \rule[0mm]{0mm}{ 3mm}{Test result}................................................................................................ &   3.92      &n.a.\\
        {Significance}............................................................................................. &  $1.26\,\sigma$&n.a.\\
        {$ 2\sigma$ U.~L.~on excess events}....................................................................... & 1635 &3198\\
        {U.~L.~on the Cutoff energy (GeV)}........................................................... &27 &34\\
        {$ 2\sigma$ Integral flux limit above 60\,GeV (cm$^{-2}\,$s$^{-1}$)......................................} & $2.5\cdot10^{-11}$&$7.9\cdot10^{-11}$\\
        {$ 2\sigma$ Differential flux limit at 60\,GeV (cm$^{-2}\,$s$^{-1}\,$GeV$^{-1}$) }..........................&$4.5\cdot10^{-12}$&$8.9\cdot10^{-12}$\\
        \rule[-2mm]{0mm}{ 0mm}{Peak Energy MC (GeV)}..........................................................................&{60}&60\\\tableline
       \end{tabular}
    \end{table*}

The spectrum with a given $E_\mathrm{Cutoff}$ is convoluted with
the effective collection area after cuts. The collection area is
derived from MC simulations, assuming the same $\gamma$-ray
spectrum. The number of expected excess events for the assumed
cutoff energy is obtained by multiplying the convoluted spectrum
with the observation time. In an iterative algorithm
$E_{\mathrm{Cutoff}}$ is changed until the number of expected
excess events matches the upper limit on the number of excess
events. In this way we derive an upper limit on the cutoff energy
of  $27\,$GeV from the result of the \emph{H}-test and 34\,GeV from the
limit obtained from the pulse phase profile.
Differential and integral upper limits on the flux were calculated
and are shown in Table \ref{crab300phe}.

\section{Discussion}\label{discussion}

In this paper we report on the most detailed study to date of VHE
$\gamma$-ray emission of the Crab Nebula below 500\,GeV. This
study includes the following:
   \begin{itemize}
      \item{A measurement of the differential energy spectrum down to 60\,GeV.}
      \item{An estimate of the peak in the SED of the VHE $\gamma$-ray emission.}
      \item{The search for an extended source morphology.}
      \item{The calculation of a light curve of the VHE $\gamma$-ray emission from the nebula above 200\,GeV.}
      \item{A search for pulsed emission from the Crab pulsar in differential bins of energy and in an optimized low-energy analysis.}
   \end{itemize}
Most of the aforementioned studies were done in this energy region
for the first time; they were possible only because the imaging
air shower Cerenkov technique was used. The wave front sampling
technique, which, up to now, was the only experimental technique
used in this energy domain, did, at best, allow one to arrange an
integral flux measurement and to search for pulsed emission. The
performance of MAGIC is superior even in the energy domain below
$200\,$GeV, where a progressive degradation of the
$\gamma$/hadron-separation power is observed. In this context
studies for improving the suppression of background events by
exploiting the intrinsic time structure of the recorded
PMT-signals are ongoing. Further improvement is expected with the
second MAGIC telescope currently under construction.

The measured energy spectrum of the Crab Nebula in Figure
\ref{crab:diffspectr} extends over two decades in energy and five
decades in flux. The spectral shape deviates from a pure power-law
behavior and is, within experimental uncertainties, in agreement
with a curved power-law. The observation supports the generally
accepted picture that the steady emission above a few tens of GeV
and up to the highest measured energies can be described within
the framework of the SSC-model
\citep{1965PhRvL..15..577G,1992ApJ...396..161D,2004ApJ...614..897A}.
Also the peak position of the inverse Compton emission of the
tested predictions is in agreement with the estimated peak in the
SED  ($77\pm47_\mathrm{stat}{+107\atop
-46}_\mathrm{syst}$\,GeV).

At GeV energies EGRET observed  a $\gamma$-ray flux, which was a
factor of 5 above the flux predicted by the SSC-mechanism
\citep{1996ApJ...457..253D}. \cite{1996MNRAS.278..525A} explain
this $\gamma$-ray excess by an additional $\gamma$-ray component
from bremsstrahlung of electrons that are partially captured in
filaments of the nebula. Such an extra component can significantly
change the spectral slope at several hundred GeV compared to a
pure IC-scenario (cf.~\emph{blue and red lines} in Figure
\ref{photindex}) and results in an almost pure power-law behavior
of the energy spectrum between $\sim100\,$GeV and $10\,$TeV
(constant $\Gamma$). At several hundred GeV, where the measurement
is most sensitive, the measured slope (\emph{black line and data point})
is considerably harder than predicted by
\cite{1996MNRAS.278..525A}. It is, therefore, unlikely that the
$\gamma$-ray excess at GeV energies can be explained by
bremsstrahlung as proposed. Later predictions by
\cite{1998nspt.conf..439A} that also include the mentioned
bremsstrahlung mechanism, are in agreement with the presented
measurement (\emph{red line}). However, all above-mentioned predictions
agree with the measurement if the measured slope is shifted by 0.2
to more negative values, which is within the range of the
systematic uncertainty of the measurement. In the prediction by
\cite{1998nspt.conf..439A} also a $\gamma$-ray component from
$\pi^0$-decay is included, which results in a considerable harder
spectrum above a few TeV compared to the pure IC-scenario
(cf.~Figure \ref{photindex}). However, given the limited
statistics above 1\,TeV of our measurement, one cannot exclude any
such prediction from the measurement.

Studies about the morphology of the $\gamma$-ray-emitting region
of the Crab Nebula have been performed by
\cite{2000A&A...361.1073A,2004ApJ...614..897A} for $\gamma$-ray
energies above $1\,$TeV. In both cases it was found that within
the resolution of the experiment the emission region is pointlike.
They placed an upper limit on the source size of $\sim2\arcmin$ at
energies between 1 and 10\,TeV. In the VHE domain, the morphology
of the emission region has not yet been studied at energies below
1\,TeV. With the resolution of MAGIC it was possible to constrain
the origin of the $\gamma$-ray emission to be within the optical
synchrotron nebula (see Figure \ref{crabcomposite}). The upper
limit on the size of the emission region is $\sim2\arcmin$, which
is about 4 times larger than the predicted size of the inverse
Compton surface brightness for $\gamma$-ray energies below
500\,GeV \citep{1992ApJ...396..161D}.

X-ray observations indicate variabilities in the acceleration and
cooling times of electrons on timescale of  months
\citep[e.g.][]{2002ApJ...577L..49H}. However, variations in
$\gamma$-rays could not be detected so far. The sensitivity of
MAGIC allowed us to study the variability of the $\gamma$-ray
emission above $150\,$GeV on timescales as short as a few minutes
up to months. We measured a flux that is within statistics
compatible with steady emission. During the observation the
stability of the integral flux was better than 10\% on all tested
timescales.

In a search for pulsed VHE $\gamma$-ray emission with MC optimized
cuts in HADRONNESS and ALPHA an excess was found
in the pulse phase profile at the same position where EGRET
detected pulsed emission above 100\,MeV. The significance of the
excess is $2.9\,\sigma$ if the phase regions where EGRET detected
pulsed emission were chosen as signal regions and the remaining
phase intervals are considered as background regions. The
similarity of the distribution of excess events in the EGRET
$>5\,$GeV and MAGIC data and the monotonic increase of the number
of excess events with increasing upper SIZE cut are
strong indications that the observed excess is not a random
fluctuation.

With the result of the \emph{H}-test an upper limit on the cutoff energy
of 27\,GeV was derived, assuming that the power-law spectrum of
the pulsar at GeV energies is attenuated by an exponential cutoff.
However, if the cutoff of the spectrum has a super-exponential
shape, a cutoff energy almost as high as the analysis threshold
($\sim60\,$GeV) cannot be excluded.

With the derived upper limit we constrain not only the
$\gamma$-ray emission from within the light cylinder but also the
predicted pulsed $\gamma$-ray emission in the unshocked wind
region \citep{2000MNRAS.313..504B}. The predicted $\gamma$-ray
flux strongly depends (1) on the distance from the light cylinder
where the kinetic energy-dominated wind forms and (2) on the
wind's Lorentz factor. By comparing our observational limits on
the pulsed emission with the predicted spectra by
\cite{2000MNRAS.313..504B}, we can exclude the formation of a
particle-dominated wind within a few light cylinder radii.
Following the argumentation of the same authors, the particle-dominated pulsar wind must therefore be formed farther out, most
likely at distances of more than 30 light cylinder radii.

Also, no pulsed emission was detected for energies above 100 GeV,
which could have its origin in IC-upscattering of IR photons
within the light cylinder. Despite earlier claims of a strong
component \citep{2001ApJ...558..216H}, latest models
\citep{2007astro.ph..1676H} seem to disfavor a pulsed TeV
component from the Crab pulsar due to dominant $\gamma$-$\gamma$
absorption processes. In the future, detailed spectroscopic studies of
the pulsed emission by, e.g., \emph{GLAST} and ground-based experiments
with lower thresholds and higher sensitivities like MAGIC II
(under construction) and CTA (projected) will hopefully resolve
the long-standing question of the origin of the pulsed emission.

\acknowledgements

We are grateful for discussions with Kouichi Hirotani. We also
would like to thank the IAC for the excellent working conditions
at the ORM in La Palma. The support of the German BMBF and MPG,
the Italian INFN, the Spanish CICYT, ETH research grant TH 34/04
3, and the Polish MNiI grant 1P03D01028 is gratefully
acknowledged.

\bibliographystyle{plainnat}
%\bibliography{pulsars,crab,others,gamma_analysis}
\bibliography{ms}

\begin{thebibliography}{83}
\providecommand{\natexlab}[1]{#1}
\providecommand{\url}[1]{\texttt{#1}}
\expandafter\ifx\csname urlstyle\endcsname\relax
  \providecommand{\doi}[1]{doi: #1}\else
  \providecommand{\doi}{doi: \begingroup \urlstyle{rm}\Url}\fi

\bibitem[{Acharya} et~al.(1992){Acharya}, {Bhat}, {Gandhi}, {Ramana Murthy},
  {Sathyanarayana}, and {Vishwanath}]{1992A&A...258..412A}
 {Acharya}, B.~S.,  {Bhat}, P.~N., {Gandhi},  V.~N.,  {Ramana Murthy}, P.~V., 
  {Sathyanarayana}, G.~P., \& {Vishwanath}, P.~R. 1992,
\newblock {\aap}, 258, 412

\bibitem[{Aharonian} et~al.(1999){Aharonian},  and
  Others]{1999A&A...346..913A}
{Aharonian}, F. et~al. 1999, \newblock {\aap}, 346, 913


\bibitem[{Aharonian} et~al.(2004){Aharonian}, and Others]{2004ApJ...614..897A}
---------. 2004,
  \newblock {\apj}, 614, 897


\bibitem[{Aharonian} et~al.(2006){Aharonian},  and Others]{2006A&A...457..899A}
---------. 2006,
\newblock {\aap}, 457, 899

\bibitem[{Aharonian} and {Atoyan}(1998)]{1998nspt.conf..439A}
 {Aharonian}, F.~A., \& {Atoyan}, A.~M. 1998,
\newblock  {in Neutron Stars and Pulsars: Thirty
  Years after the Discovery, ed. N.~{Shibazaki} (Tokyo: Universal Academy)}, 439



\bibitem[{Aharonian} et~al.(2000){Aharonian},  and Others]{2000A&A...361.1073A}
 {Aharonian}, F.~A., et~al. 2000,
\newblock {\aap}, 361, 1073

\bibitem[{Akerlof} et~al.(1990){Akerlof}, and Others]{1990ICRC....2..135A}
{Akerlof}, C., {Dimarco}, J., {Levy}, H., {MacCallum}, C., {Meyer}, D.,
  {Radusewicz}, P., {Tschirhart}, R., \& {Yama},  Z. 1990,
\newblock in { Proc, 21st Int. Cosmic Ray Conf. (Adelaide)}, 135



\bibitem[{Albert} et~al.(2006){Albert}, and Others]{magicextraction}
{Albert}, J., et~al. 2006, preprint (astro-ph/0612385)


\bibitem[{Albert} et~al., 2007a]{albert:2007:magic:mkn421}
---------. 2007a,
\newblock {\em \apj}, 663, 125

\bibitem[{Albert et al.}(2007b){Albert}, and Others]{magicrandomforest}
---------. 2007b, preprint (astro-ph/0709.3719)

\bibitem[{Albert et al.}(2007c){Albert}, and Others]{magicunfolding}
---------. 2007c,
 {Nucl.Instrum. Methods Phys. Res. A},
  583, 494


\bibitem[{Amato} et~al.(2003){Amato}, {Guetta}, and
  {Blasi}]{2003A&A...402..827A}
{Amato}, E., {Guetta}, D., \&~{Blasi}, P. 2003,
\newblock {\aap}, 402, 827

\bibitem[{Amenomori} et~al.(1999){Amenomori}, and Others {The Tibet As
  {$\Gamma$} Collaboration}]{1999ApJ...525L..93A}
{Amenomori}, M., et~al. 1999,
\newblock {\apjl}, 525, L93

\bibitem[{Anykeyev} et~al.(1991){Anykeyev}, {Spiridonov}, and
  {Zhigunov}]{1991NIMPA.303..350A}
 {Anykeyev}, V.~B., {Spiridonov},  A.~A., \& {Zhigunov},  V.~P. 1991,
Nucl.~Instrum.~Meth.~Phys.~Res.~A,
  303, 350

\bibitem[{Arons}(1983)]{1983ApJ...266..215A}
{Arons}, J. 1983,
\newblock {\apj}, 266, 215

\bibitem[{Arqueros} et~al.(2002){Arqueros},  and
  Others]{2002APh....17..293A}
{Arqueros}, F., et~al. 2002,
\newblock {Astropart. Phys.}, 17, 293

\bibitem[{Atoyan} and {Aharonian}(1996)]{1996MNRAS.278..525A}
{Atoyan}, A.~M., \& {Aharonian},  F.~A. 1996,
\newblock {\mnras}, 278, 525

\bibitem[{Baillon} et~al.(1991)]{1991ICRC....1..220B}
{Baillon}, P., et~al. 1991,
\newblock in Proc. 22nd {Int. Cosmic Ray Conf. (Dublin)}, 220

\bibitem[{Bednarek} and {Bartosik}(2003)]{2003A&A...405..689B}
{Bednarek}, W., \& {Bartosik},  M. 2003,
\newblock {\aap}, 405, 689

\bibitem[{Bednarek} and {Protheroe}(1997)]{1997PhRvL..79.2616B}
{Bednarek}, W., \&  {Protheroe}, R.~J. 1997, 
\newblock {Phys.~Rev.~Lett.}, 79, 2616

\bibitem[Bertero(1989)]{Bertero88}
Bertero, M. 1989,
\newblock {Electronics and Electron Phys.}, 75, 1989.

\bibitem[{Bhat} et~al.(1986){Bhat}, {Ramanamurthy}, {Sreekantan}, and
  {Vishwanath}]{1986Natur.319..127B}
 {Bhat}, P.~N.,  {Ramanamurthy}, P.~V.,  {Sreekantan}, B.~V., \& {Vishwanath}, P.~R. 1986,
\newblock {\nat}, 319, 127

\bibitem[{Bock} et~al.(2004){Bock},  and Others]{2004NIMPA.516..511B}
 {Bock}, R.~K., et~al. 2004,
\newblock {Nucl.~Instrum.~Methods Phys.~Res.~A},
  516, 511

\bibitem[{Bogovalov} and {Aharonian}(2000)]{2000MNRAS.313..504B}
 {Bogovalov}, S.~V., \& {Aharonian}, F.~A. 2000,
\newblock {\mnras}, 313, 504

\bibitem[{Breiman}(2001)]{2001breiman}
{Breiman}, L. 2001,
\newblock {Machine Learning}, 45, 5

\bibitem[{Bretz} and {Wagner}(2003)]{2003ICRCMARS}
{Bretz} T., \&  {Wagner}, R.~M. 2003,
\newblock in Proc. 28th {Int.~Cosmic Ray Conf.~Tsukuba}, 2947

\bibitem[{Cheng} et~al.(1986{\natexlab{a}}){Cheng}, {Ho}, and
  {Ruderman}]{1986ApJ...300..500C}
 {Cheng}, K.~S., {Ho},  C., \& {Ruderman},  M. 1986a,
\newblock {\apj}, 300, 500

\bibitem[{Cheng} et~al.(1986{\natexlab{b}}){Cheng}, {Ho}, and
  {Ruderman}]{1986ApJ...300..522C}
---------. 1986b,
\newblock {\apj}, 300, 522

\bibitem[{Chiang} and {Romani}(1992)]{1992ApJ...400..629C}
{Chiang}, J. \& {Romani},  R.~W. 1992,
\newblock {\apj}, 400, 629

\bibitem[{Cocke} et~al.(1969){Cocke}, {Disney}, and
  {Taylor}]{1969Natur.221..525C}
 {Cocke}, W.~J.,  {Disney}, M.~J., \&  {Taylor}, D.~J. 1969,
\newblock {\nat}, 221, 525

\bibitem[{Collins} et~al.(1999){Collins}, {Claspy}, and
  {Martin}]{1999PASP..111..871C}
 {Collins}, G.~W., II,  {Claspy}, W.~P., \& {Martin},  J.~C. 1999,
\newblock {\pasp}, 111, 871

\bibitem[{Cortina} et~al.(2005)]{2005ICRC....5..359C}
{Cortina}, J.,  et~al. 2005,
\newblock in Proc. 29th {Int.~Cosmic Ray Conf.~(Pune)}, 359

\bibitem[{Daugherty} and {Harding}(1982)]{1982ApJ...252..337D}
 {Daugherty}, J.~K., \&  {Harding}, A.~K. 1982,
\newblock {\apj}, 252, 337

\bibitem[{de Jager}(1994)]{1994ApJ...436..239D}
 {de Jager}, O.~C. 1994,
\newblock {\apj}, 436, 239

\bibitem[{de Jager} and {Harding}(1992)]{1992ApJ...396..161D}
 {de Jager}, O.~C., \&  {Harding}, A.~K. 1992,
\newblock {\apj}, 396, 161

\bibitem[{de Jager} et~al.(1996){de Jager}, and Others]{1996ApJ...457..253D}
 {de Jager}, O.~C.,  {Harding}, A.~K.,  {Michelson}, P.~F.,  {Nel}, H.~I., 
  {Nolan}, P.~L., {Sreekumar}, P.~, \&  {Thompson}, D.~J.
\newblock {\apj}, 457, 253

\bibitem[{de Jager} et~al.(1989){de Jager}, {Raubenheimer}, and
  {Swanepoel}]{1989A&A...221..180D}
 {de Jager}, O.~C., {Raubenheimer}, B.~C., \&  {Swanepoel}, J.~W.~H. 1989,
\newblock {\aap}, 221, 180

\bibitem[{de Naurois} et~al.(2002){de Naurois},  and Others]{2002ApJ...566..343D}
{de Naurois}, M., et~al., 2002
\newblock {\apj}, 566, 343

\bibitem[{Domingo-Santamar\'\i a} et~al.(2005)]{2005ICRC....5..363D}
{Domingo-Santamar\'\i a}, E., et~al. 2005,
\newblock in Proc. 29th {Int.~Cosmic Ray Conf.~(Pune)}, 363

\bibitem[{Downthwaite} et~al.(1984){Downthwaite}, {Harrison}, {Kirkman},
  {Macrae}, {McComb}, {Orford}, {Turver}, and {Walmsley}]{1984ApJ...286L..35D}
 {Downthwaite}, J.~C.,  {Harrison}, A.~B. {Kirkman}, I.~W., {Macrae},
H.~J., {McComb},   T.~J.~L., {Orford}, K.~J., {Turver}, K.~E.,\& {Walmsley},M.~1984,
\newblock {\apjl}, 286, L35

\bibitem[{Fierro} et~al.(1998){Fierro}, {Michelson}, {Nolan}, and
  {Thompson}]{1998ApJ...494..734F}
 {Fierro}, J.~M., {Michelson}, P.~F., {Nolan}, P.~L., \&  {Thompson}, D.~J. 1998
\newblock {\apj}, 494, 734

\bibitem[{Fomin} et~al.(1994){Fomin}, {Stepanian}, {Lamb}, {Lewis}, {Punch},
  and {Weekes}]{1994APh.....2..137F}
 {Fomin}, V.~P. {Stepanian}, A.~A., {Lamb}, R.~C., {Lewis}, D.~A.,~{Punch}, M. \&
   {Weekes}; T.~C. 1994,
\newblock {Astropart.~Phys.}, 2, 137

\bibitem[{Garczarczyk}(2006)]{garc07}
{Garczarczyk}, M.~2006,
\newblock Ph.D.~thesis, Univ.~Rostock

\bibitem[{Gaug} et~al.(2005){Gaug}, {Bartko}, {Cortina}, and
  {Rico}]{magiccalibration}
{Gaug}, M.,~{Bartko}, H.,~{Cortina}, J., \& {Rico}, J. 2005,
\newblock in Proc. 29th Int.~Cosmics
  Rays Conf.~(Pune), 375

\bibitem[{Gibson} et~al.(1982){Gibson}, {Harrison}, {Kirkman}, {Lotts},
  {Macrae}, {Orford}, {Turver}, and {Walmsley}]{1982Natur.296..833G}
 {Gibson}, A.~I., {Harrison}, A.~B., {Kirkman}, I.~W., {Lotts}, A.~P.,
  {Macrae}, J.~H., {Orford}, K.~J., {Turver}, K.~E., \& {Walmsley}, M. 1982,
\newblock {\nat}, 296, 833

\bibitem[{Goret} et~al.(1993){Goret}, {Palfrey}, {Tabary}, {Vacanti}, and
  {Bazer-Bachi}]{1993A&A...270..401G}
{Goret}, P.,~{Palfrey}, T.,~{Tabary}, A.,~{Vacanti}, G., \& {Bazer-Bachi}, R.~1993
\newblock {\aap}, 270, 401

\bibitem[{Gould}(1965)]{1965PhRvL..15..577G}
 {Gould}, R.~J. 1965,
\newblock {Phys.~Rev.~Lett.}, 15, 577

\bibitem[{Gregory} and {Loredo}(1992)]{1992ApJ...398..146G}
 {Gregory}, P.~C., \&  {Loredo}, T.~J. 1992,
\newblock {\apj}, 398, 146

\bibitem[{Harding} et~al.(1978){Harding}, {Tademaru}, and
  {Esposito}]{1978ApJ...225..226H}
 {Harding}, A.~K.,~{Tademaru}, E., \&  {Esposito}, L.~W. 1978,
\newblock {\apj}, 225, 226

\bibitem[{Hester} et~al.(2002){Hester}, {Mori}, {Burrows}, {Gallagher},
  {Graham}, {Halverson}, {Kader}, {Michel}, and {Scowen}]{2002ApJ...577L..49H}
 {Hester}, J.~J., et al. 2002,
\newblock {\apjl}, 577, L49

\bibitem[{Hillas}(1985)]{1985ICRC....3..445H}
 {Hillas}, A.~M. 1985, 
\newblock in Proc.~19th Int.~Cosmic Ray Conf.~(La Lolla), 445

\bibitem[{Hillas} et~al.(1998){Hillas}, and
  Others]{1998ApJ...503..744H}
 {Hillas}, A.~M., et~al. 1998,
\newblock {\apj}, 503, 744

\bibitem[{Hirotani}(2001)]{2001ApJ...549..495H}
{Hirotani}, K.~2001,
\newblock {\apj}, 549, 495

\bibitem[{Hirotani}(2007)]{2007astro.ph..1676H}
---------. 2007,
\newblock {\apj}, 662, 1173.

\bibitem[{Hirotani} and {Shibata}(2001)]{2001ApJ...558..216H}
{Hirotani}, K., \&~{Shibata}, S. 2001,
\newblock {\apj}, 558, 216

\bibitem[{Kennel} and {Coroniti}(1984{\natexlab{a}})]{1984ApJ...283..694K}
 {Kennel}, C.~F., \& {Coroniti}, F.~V. 1984a,
\newblock {\apj}, 283, 694

\bibitem[{Kennel} and {Coroniti}(1984{\natexlab{b}})]{1984ApJ...283..710K}
---------. 1984b,
\newblock {\apj}, 283, 710

\bibitem[{Konopelko} et~al.(1996){Konopelko},  and Others]{1996APh.....4..199H}
{Konopelko}, A., et~al. 1996,
\newblock {Astropart.~Phys.}, 4, 199

\bibitem[{Lessard} et~al.(2000){Lessard},  and Others]{2000ApJ...531..942L}
 {Lessard}, R.~W., et~al. 2000,
\newblock {\apj}, 531, 942

\bibitem[{Lessard} et~al.(2001){Lessard}, {Buckley}, {Connaughton}, and {Le
  Bohec}]{2001APh....15....1L}
{Lessard}, R.~W., {Buckley}, J.~H.,~{Connaughton}, V., \&~{Le Bohec}, S. 2001,
\newblock {Astropart.~Phys.}, 15, 1

\bibitem[{Li} and {Ma}(1983)]{1983ApJ...272..317L}
 {Li}, T.-P., \& {Ma}, Y.-Q. 1983,
\newblock {\apj}, 272, 317

\bibitem[{Lorenz}(2004)]{2004NewAR..48..339L}
{Lorenz}, E. 2004,
\newblock {NewA.~Rev.}, 48, 339

\bibitem[{Lucarelli} et~al.(2005)]{2005ICRC....5..367L}
{Lucarelli}, F., et~al. 2005,
\newblock in Proc. 29th {Int.~Cosmic Ray Conf.~(Pune)}, 367

\bibitem[{Majumdar} et~al.(2002){Majumdar},  and Others]{2002majumdar}
{Majumdar}, P., et~al. 2002,
\newblock in {{The Universe Viewed in Gamma-Rays, ed.~R.~Enomoto, M.~Mori, \& S.~Yanagita (Tokyo: Universal Academy)}}

\bibitem[Mirzoyan and Lorenz(1997)]{ffactororig}
Mirzoyan, R., \& Lorenz, E. 1997,
\newblock in Proc.~25th {{Int.~Cosmic Ray
  Conf. (Durban)}} 265

\bibitem[{Muslimov} and {Harding}(2003)]{2003ApJ...588..430M}
 {Muslimov} A.~G., \& {Harding}, A.~K. 2003
\newblock {\apj}, 588, 430

\bibitem[{Musquere}(1999)]{1999ICRC....3..460M}
{Musquere} A.~1999,
\newblock in Proc.~26th {{Int.~Cosmic Ray
  Conf. (Salt Lake City)}}, 460

\bibitem[{Nolan} et~al.(1993){Nolan}, 
  and Others]{1993ApJ...409..697N}
 {Nolan}, P.~L., et~al. 1993
\newblock {\apj}, 409, 697

\bibitem[{Oosterbroek} et~al.(2006){Oosterbroek}, {de Bruijne}, {Martin},
  {Verhoeve}, {Perryman}, {Erd}, and {Schulz}]{2006astro.ph..6146O}
{Oosterbroek}, T., {de Bruijne}, J.~H.~J.,~{Martin}, D.,~{Verhoeve}, P.,
  {Perryman}, M.~A.~C.,~{Erd}, C., \&~{Schulz}, R. 2006, A\&A, 456, 283

\bibitem[{Oser} et~al.(2001){Oser}, 
  and Others]{2001ApJ...547..949O}
{Oser}, S., et~al. 2001,
\newblock {\apj}, 547, 949

\bibitem[{Paneque} et~al.(2003){Paneque}, {Gebauer}, {Lorenz}, {Martinez},
  {Mase}, {Mirzoyan}, {Ostankov}, and {Schweizer}]{2003NIMPA.504..109P}
{Paneque}, D., {Gebauer}, H.~J.,~{Lorenz}, E.,~{Martinez}, M.~{Mase},
  K.,~{Mirzoyan}, R.,~{Ostankov}, A., \&~{Schweizer}, T. 2003,
\newblock {Nucl.~Instrum.~Methods Phys.~Res.~A},
  504, 109

\bibitem[{Rees} and {Gunn}(1974)]{1974MNRAS.167....1R}
 {Rees}, M.~J., \& {Gunn}, J.~E. 1974,
\newblock {\mnras}, 167, 1

\bibitem[{Rolke} et~al.(2005){Rolke}, {L{\'o}pez}, and
  {Conrad}]{2005NIMPA.551..493R}
{Rolke}, W.~A., {L{\'o}pez}, A.~M., \&~{Conrad}, J. 2005,
\newblock {Nucl.~Instrum.~Methods Phys.~Res.~A},
  551, 493

\bibitem[{Schmelling}(1994)]{1994NIMPA.340..400S}
{Schmelling}, M. 1994,
\newblock {Nucl.~Instrum.~Methods Phys.~Res.~A},
  340, 400

\bibitem[{Smith} et~al.(2006)]{2006A&A...459..453S}
 {Smith}, D.~A., et~al. 2006, 
\newblock {\aap}, 459, 453


\bibitem[{Staelin} and {Reifenstein}(1968)]{1968Sci...162.1481S}
 {Staelin}, D.~H., \& {Reifenstein}, E.~C. 1968,
\newblock {Science}, 162, 1481

\bibitem[{Tanimori} et~al.(1998){Tanimori},  and
  Others]{1998ApJ...492L..33T}
{Tanimori}, T., et~al. 1998,
\newblock {\apjl}, 492, L33

\bibitem[{Taylor} et~al.(2000){Taylor}, {Manchester}, {Nice}, {Weisberg}, A.,
  and N.]{tempo}
 {Taylor}, J.~H., {Manchester}, R.~N., {Nice}, D.~J., {Weisberg}, J.~M., {Irwin} A.,
  \& {Wex} N. 2000,
\newblock Tempo Pulsar Timing Package

\bibitem[{Thompson} et~al.(1999){Thompson},  and
  Others]{1999ApJ...516..297T}
 {Thompson}, D.~J., et~al. 1999,
\newblock {\apj}, 516, 297

\bibitem[{Thompson} et~al.(2004){Thompson},  and
  Others]{2004ASSL..304..149T}
---------. 2004,
\newblock in Cosmic Gamma-Ray Sources, ed.~K.~S.~Cheng \& G.~E.~Romero (Dordrecht: Kluwer), 149

\bibitem[{Thompson} et~al.(2005){Thompson},  {Bertsch} and {O'Neal}]{2005ApJS..157..324T}
---------. 2005,
\newblock {\apjs}, 157, 324

\bibitem[Tikhonov and Arsenin(1979)]{Tikhonov79}
 Tikhonov, {A.\,N.}, \&  Arsenin, {V.\,J.} 1979,
\newblock Methods of Solutions of Ill-posed Problems {(Washington: Winston \& Sons)}

\bibitem[{Vacanti} et~al.(1991){Vacanti}, and Others]{1991ApJ...377..467V}
{Vacanti}, G., et~al. 1991,
\newblock {\apj}, 377, 467

\bibitem[{Weekes} et~al.(1989){Weekes},  and
  Others]{1989ApJ...342..379W}
 {Weekes}, T.~C., et~al. 1989,
\newblock {\apj}, 342, 379

\bibitem[{Yao} et~al.(2006){Yao}, 
  and Others]{PDBook}
 {Yao}, W.-M., et~al. 2006,
\newblock {{J.~Phys.~G}}, 33, 1
\end{thebibliography}

\clearpage
\begin{landscape}
\begin{table}[h]

\thispagestyle{empty}
    \caption{\label{syserrors} Contribution to the Systematic Uncertainties}

       \begin{tabular}{clccl}\hline\hline
       &&&Uncertainty&\\
       \rule[0mm]{0mm}{ 3mm}\rule[-2mm]{0mm}{ 0mm}{Item} & \multicolumn{1}{c}{Source of Uncertainty} & {Class} &
        {(\%)}&\multicolumn{1}{c}{Comments}\\\hline
         1........... & Parametrization of Atmosphere in MC-simulation & A & 3
         &
          \parbox[t]{10cm}{\renewcommand{\baselinestretch}{1}\footnotesize Deviations due to yearly and daily
         pressure changes, deviations of real density distribution
         and standard atmosphere model} \\
        2........... & Atmospheric transmission losses due to Mie scattering  & A, (C) & 5
         &
           Lack of good measurements; short term unpredictable changes possible \\
        3...........& \parbox{5cm}{\renewcommand{\baselinestretch}{1}\footnotesize Incorrect NSB simulation} & A &
3
         &
          \parbox[t]{9cm}{\renewcommand{\baselinestretch}{1}\footnotesize MC assumes uniform NSB. Variations due to source location, air glow, variations due to manmade light. Stars in the FoV.} \\

          4...........& \parbox{4cm}{\renewcommand{\baselinestretch}{1}\footnotesize Reflectivity of main mirror} & A & 7
         &
          \parbox{7cm}{\renewcommand{\baselinestretch}{1}\footnotesize From measurements of reflected star images} \\

          5........... & \parbox{7cm}{\renewcommand{\baselinestretch}{1}\footnotesize Variation of the useful mirror area} & A & 3
         &
          \parbox[t]{10.5cm}{\renewcommand{\baselinestretch}{1}\footnotesize Malfunctions of active mirror control resulting in focussing losses} \\

          6...........& \parbox{4cm}{\renewcommand{\baselinestretch}{1}\footnotesize Day to day reflectivity changes} & A & 2
         &
          \parbox{10.5cm}{\renewcommand{\baselinestretch}{1}\footnotesize Due to dust deposit variations and occasional dew deposit } \\
  7...........& \parbox{7.5cm}{\renewcommand{\baselinestretch}{1}\footnotesize Photon detection efficiency of the PMT/lightcatcher system} & A, C & 10-12
         &
          \parbox{7cm}{\renewcommand{\baselinestretch}{1}\footnotesize See text } \\

8...........& \parbox{7cm}{\renewcommand{\baselinestretch}{1}\footnotesize Unusable
camera channels} & B & 3
         &
          \parbox{10cm}{\renewcommand{\baselinestretch}{1}\footnotesize Dead PMTs (5-10 channels), problems in calibration (5-10 channels)} \\

9...........& \parbox{4cm}{\renewcommand{\baselinestretch}{1}\footnotesize Trigger
inefficiencies} & B, C & 4
         &
          \parbox{12.5cm}{\renewcommand{\baselinestretch}{1}\footnotesize Due to discriminator dead-time, baseline shifts/drifts, level differences trigger branch and FADC branch etc.} \\

10..........& \parbox{7cm}{\renewcommand{\baselinestretch}{1}\footnotesize Signal
drift in camera due to temperature drifts} & A, C & 2
         &
          \parbox{10cm}{\renewcommand{\baselinestretch}{1}\footnotesize Combination of PMT QE change (small), amplifier and optical transmitter drifts} \\

11..........& \parbox{7cm}{\renewcommand{\baselinestretch}{1}\footnotesize Camera
flatfielding} & A,B & 2
         &
          \parbox{7cm}{\renewcommand{\baselinestretch}{1}\footnotesize Calibration problem} \\

12..........& \parbox{7cm}{\renewcommand{\baselinestretch}{1}\footnotesize Signal
extractor} & B & 5
         &
          \parbox[t]{10cm}{\renewcommand{\baselinestretch}{1}\footnotesize Complex effect due to trigger jitter
          (early pulses from PEs generated on 1st dynode) etc.; baseline jitter, shifts in FADCs} \\

13..........& \parbox{7cm}{\renewcommand{\baselinestretch}{1}\footnotesize cuts and
methods used in the analysis} & B,C & 5-30
         &
          \parbox{10cm}{\renewcommand{\baselinestretch}{1}\footnotesize Energy dependent, see discussion of differential energy spectrum} \\

14..........& \parbox{7cm}{\renewcommand{\baselinestretch}{1}\footnotesize Losses
of events during reconstruction} & B(A) & 8
         &
          \parbox{10cm}{\renewcommand{\baselinestretch}{1}\footnotesize
          Simplifications in MC simulation} \\

15..........& \parbox{7cm}{\renewcommand{\baselinestretch}{1}\footnotesize Estimate
of BG under source} & B(A) & 4
         &
          \parbox{12cm}{\renewcommand{\baselinestretch}{1}\footnotesize Camera nonuniformity not included in MC. Hadronic events not perfectly simulated in MC.} \\

16..........& \parbox{7cm}{\renewcommand{\baselinestretch}{1}\footnotesize Small
tracking instabilities} & B & 2
         &
          \parbox[t]{11.5cm}{\renewcommand{\baselinestretch}{1}\footnotesize Source jitters around nominal camera position due to small tracking errors, small camera oscillations due to gusts etc, resulting in a wider signal spread than predicted by MC}
          \\
17..........& \parbox{7cm}{\renewcommand{\baselinestretch}{1}\footnotesize
Nonlinearities in the analog signal chain (PMT--FADC} & C(A) &
3-10
         &
          \parbox{10cm}{\renewcommand{\baselinestretch}{1}\footnotesize Saturation and nonlinearities of electronic and opto-electronic components} \\\hline

       \end{tabular}

    \tablecomments{Class A: contributions to the uncertainty on the energy scale. Class B: contributions to the uncertainty in the event rate. Class B(A): error contributes more to the leading term. Some of the uncertainties are energy dependent and are averaged. Class C: contribution affecting the spectral slope.}

    \end{table}
    \clearpage
\end{landscape}

\end{document}